\documentclass[fleqn,usenatbib]{mnras}

\usepackage[T1]{fontenc}

\DeclareRobustCommand{\VAN}[3]{#2}
\let\VANthebibliography\thebibliography
\def\thebibliography{\DeclareRobustCommand{\VAN}[3]{##3}\VANthebibliography}

\usepackage{graphicx}	
\usepackage{amsmath}	
\usepackage{amssymb}	
\usepackage{physics}
\usepackage{bm}

\renewcommand{\vec}{\bm}

\title[Eccentric Debris discs]{Eccentric debris belts reveal the dynamical history of the companion exoplanet}

\author[Laetitia Rodet \& Dong Lai]{
Laetitia Rodet,$^{1}$\thanks{E-mail:lbr63@cornell.edu}
Dong Lai,$^{1}$
\\
$^{1}$Cornell Center for Astrophysics and Planetary Science, Department of Astronomy, Cornell University, Ithaca, NY 14853, USA
}

\date{Accepted 2022 September 8. Received 2022 September 7; in original form 2022 August 10}

\pubyear{2015}

\begin{document}
\label{firstpage}
\pagerange{\pageref{firstpage}--\pageref{lastpage}}
\maketitle

\begin{abstract}
In recent years, a number of eccentric debris belts have been observed in extrasolar systems. The most common explanation for their shape is the presence of a nearby eccentric planetary companion. The gravitational perturbation from such a companion would induce periodic eccentricity variations on the planetesimals in the belt, with a range of precession frequencies. The overall expected shape is an eccentric belt with a finite minimum width. However, several observed eccentric debris discs have been found to exhibit a narrower width than the theoretical expectation. In this paper, we study two mechanisms that can produce this small width: (i) the protoplanetary disc can interact with the planet and/or the planetesimals, slowly driving the eccentricity of the former and damping the eccentricities of the latter; (ii) the companion planet could have gained its eccentricity stochastically, through planet-planet scatterings. We show that under appropriate conditions, both of these scenarios offer a plausible way to reduce the minimum width of an eccentric belt exterior to a perturbing planet. However, the effects of protoplanetary discs are diminished at large separations ($a > 10$ au) due to the scarcity of gas and the limited disc lifetime. These findings suggest that one can use the shape and width of debris discs to shed light on the evolution of extrasolar systems, constraining the protoplanetary disc properties and the prevalence of planet-planet scatterings. Further observations of debris-harbouring systems could confirm whether thin debris belts are a common occurrence, or the results of rare initial conditions or evolutionary processes.
\end{abstract}




\section{Introduction}
\label{sec:introduction}

Debris discs or belts have now been detected in over a hundred extrasolar systems. They are composed of a large variety of small bodies, with sizes ranging from micrometer (dust) to kilometer (planetesimals) \citep[see e.g.,][]{hughesDebrisDisksStructure2018}. The larger planetesimals (parent bodies) constantly replenish the population of short-lived dust through collisional cascades, which allows the disc to be detected. While small \textmu m-sized particles (that scatter light in the visible/near-infrared wavelength) can orbit far from the sites of their initial collisional formation \citep[e.g.,][]{wyattSpiralStructureWhen2005}, mm-sized particles (that emit in the millimetre wavelength) tend to remain near their production sites. They can thus be used to trace the orbital architecture of the parent bodies. 

In the Solar System, the location, density and dynamical pattern of the asteroid and Edgeworth-Kuiper belts have long served as a useful diagnosis to constrain the formation and evolutionary history of the planets. The current structure of these debris belts has largely been shaped by their interaction with the surrounding planets, not only through the violent clearing of the orbital paths, but also through more subtle resonant and secular processes. Similarly, we expect that the structure of extrasolar debris belts can help us infer the presence of planetary perturbers and their dynamical history \citep[e.g.,][]{raymondDebrisDisksSignposts2011,pearcePlanetPopulationsInferred2022,guoPlanetesimalDynamicsPresence2022}.

In the past decades, a number of debris discs have been observed to exhibit an overall eccentric shape \citep[e.g.,][]{telescoDeep10182000,kalasPlanetarySystemOrigin2005, wyattSpiralStructureWhen2005,eiroaColdDUstNEarby2010,kristHubbleSpaceTelescope2012,macgregorALMAImagesEccentric2022}. The most common explanation for this breaking of the circular symmetry is the presence of an eccentric planet in the neighbourhood of the disc \citep{wyattHowObservationsCircumstellar1999}. This planetary perturber defines a preferred direction characterized by its longitude of periastron, and induces a so-called forced eccentricity $e_{\rm forced}$ (proportional to the planetary eccentricity) in the planetesimal disc. If the planetesimals have initially circular orbits with semi-major axis $a$, then the eccentric disc would be apsidally aligned with the planet with a width of at least $\Delta a = 2 a e_{\rm forced}$ (e.g., \citealt{kennedyUnexpectedNarrownessEccentric2020}; see Section 2.2); the width can be larger if the debris belt has a spread in semi-major axis. However, several observed eccentric debris discs have been found to challenge this simple picture. In particular, at least three systems whose millimetre emissions have been observed by ALMA contain a debris belt with width narrower than the theoretical minimum \citep[as first noted by][]{kennedyUnexpectedNarrownessEccentric2020}: (i) HD 202628 has a $22$ au-wide debris belt located at $150$ au from its G-type host star \citep{faramazScatteredlightMillimeterEmission2019}. The belt has an eccentricity $0.09$, with the corresponding theoretical minimal width of $27$ au, larger than the observed width \citep[22 au,][]{kennedyUnexpectedNarrownessEccentric2020}. (ii) Fomalhaut is a well-known system comprising a massive A star and a narrow eccentric ring at $142$ au \citep{macgregorCompleteALMAMap2017,kennedyUnexpectedNarrownessEccentric2020}. This ring has an eccentricity $0.12$ and a width of $12$ au, less than half the theoretical minimum of $34$ au. This system may harbour a planetary candidate \citep{kalasOpticalImagesExosolar2008}, although recent observations suggest it is a dispersing cloud of dust, perhaps due to a collision between unseen planetary companions \citep{gasparNewHSTData2020}. (iii) HD 53143 has the most eccentric debris disc observed to date, with an eccentricity of $0.21\pm 0.02$ \citep{macgregorALMAImagesEccentric2022}. It orbits a Solar-type star at $90$ au, and has a width of $19.7 \pm 2.5$ au, about half the theoretical minimum width of $38 \pm 4$ au. All three systems are at least several hundreds Myr old.

Several hypotheses could account for the small widths of these long-lived debris belts. First, the embedded planetesimals could have finite primordial eccentricities. To reduce the width of the belt however, the primordial eccentricity vector directions should be clustered around that of the planet---such primordial alignment is not expected to occur naturally. Second, a gas-induced damping force could act to push the planetesimal eccentricity vectors toward their equilibrium values \citep[e.g.,][]{panSelfconsistentSizeVelocity2012,linSculptingEccentricDebris2019}. Since most of the gas in the system is gone after the dissipation of the protoplanetary disc, we expect the damping force to act only in the very beginning of the system's life. This requires the planet to have acquired its mass and eccentricity at a sufficiently early time. Alternatively, the damping could be induced by collisions between planetesimals in the disc \citep{kennedyUnexpectedNarrownessEccentric2020}. This effect would vary with the sizes of the planetesimals, so that we would observe different belt widths in different wavelengths. Finally, other hypotheses have been suggested that do not involve a single planet: the disc eccentricity pattern could be due to self-stirring or to a single disrupting event (such as a collision between large planetesimals), or the disc could be shaped by several shepherding planets \citep{kennedyUnexpectedNarrownessEccentric2020}.

In this paper, we explore several new promising pathways to account for the small width of eccentric debris belts using a single planetary perturber. In Section~\ref{sec:exploration}, we lay down the theoretical framework for the debris belt eccentricity evolution assuming an evolving eccentric planet. In the following sections, we then study how this framework applies to the joint evolution of ``planet + debris disc'' systems. In Section~\ref{sec:ppdisc}, we consider the effect of the protoplanetary disc, and show that the gas-induced  eccentricity excitation of the planet and the eccentricity damping of the planetesimals can both reduce the width of the final debris belt. In Section~\ref{sec:ppscattering}, we consider the scenario in which the planet eccentricity grows stochastically due to planet-planet scatterings, and show that this process has a significant probability of leading to a narrow debris belt when the scattering stops. We summarize our findings in Section~\ref{sec:conclusion}, and discuss their limitations.

\section{Test Mass perturbed by an Evolving Eccentric Planet}
\label{sec:exploration}

In this section, we summarize the key equations that govern the eccentricity evolution of a test mass under the influence of an eccentric planet, allowing the planet's mass $m_{\rm p}$ and eccentricity $e_{\rm p}$ to change in time. We assume that the orbital migration of the planet is negligible, so that the semi-major axis ratio between the planet and planetesimals remains constant.

Let $a_{\rm p}$ and $a$ be the semi-major axes of the planet and test particle. Throughout this paper, we consider debris discs exterior to the planet, thus $a > a_{\rm p}$. Let $e$ and $\varpi$ be the eccentricity and longitude of periastron of the test mass. We define the complex eccentricity
\begin{equation}
	\mathcal{E}(t) \equiv e(\cos\varpi + i \sin\varpi),
\end{equation}
and similarly $\mathcal{E}_{\rm p}(t) \equiv e_{\rm p} \exp(i\varpi_{\rm p})$ for the planet. 

For small $e_{\rm p}$ and $e$, the secular equation governing the evolution of the complex eccentricity of the test particle is
\begin{equation}
	\dv{\mathcal{E}}{t}\,(t) = i\omega(t)\mathcal{E}(t) - i \nu(t) \mathcal{E}_{\rm p}(t), \label{eq:eqdiff}
\end{equation}
where $\omega$ and $\nu$ are frequencies given by
\begin{align}
	&\omega(t) = \frac{1}{4} \mu_{\rm p}(t) \alpha b_\frac{3}{2}^{(1)}(\alpha) n = \frac{3}{4} \alpha^2  \mu_{\rm p}(t)  n + O(\alpha^4),\label{eq:frequency}\\
	&\nu(t) = \frac{1}{4} \mu_{\rm p}(t) \alpha b_\frac{3}{2}^{(2)}(\alpha) n = \frac{15}{16} \alpha^3 \mu_{\rm p}(t) n + O(\alpha^4).\label{eq:nu}
\end{align}
Here $\mu_{\rm p} = m_{\rm p}/M_*$ ($M_*$ is the mass of the host star), $n$ is the mean-motion of the test mass, $\alpha = a_{\rm p}/a$ and $b_i^{(j)}(\alpha)$ are the Laplace coefficients \citep{murraySolarSystemDynamics2000}. The second equalities in equations~\eqref{eq:frequency}--\eqref{eq:nu} are valid for $\alpha \ll 1$.

Equation \eqref{eq:eqdiff} has the formal solution:
\begin{equation}
	\mathcal{E}(t) = \left( \int_{0}^t \left[-i\nu(t')\right] \mathcal{E}_{\rm p}(t') \exp\left[-i\mathcal{W}(t')\right]  \dd{t'} + \mathcal{E}_0 \right) \exp \left[i\mathcal{W}(t)\right], \label{eq:solution}
\end{equation}
where $\mathcal{E}_0 = \mathcal{E}(t=0)$ and
\begin{equation}
	\mathcal{W}(t) \equiv \int_{0}^t \omega(t') \dd{t'}. \label{eq:W}
\end{equation}

\subsection{Constant planetary mass and eccentricity}
\label{sec:classical} 

Suppose the planetary mass and eccentricity grow instantly at $t=0$ and remain constant ($=  \mathcal{E}_{\rm p}$) thereafter. Equation~\eqref{eq:solution} then gives
\begin{equation}
	\mathcal{E}(t) = \frac{\nu}{\omega} \mathcal{E}_{\rm p} + \left(\mathcal{E}_0 - \frac{\nu}{\omega} \mathcal{E}_{\rm p} \right) \exp(i\omega t). \label{eq:classical}
\end{equation}
We see that the test mass eccentricity $\mathcal{E}(t)$ follows a circular trajectory in the complex plane. This trajectory's centre is called the ``forced eccentricity'' $e_{\rm forced}$ and its radius the ``free eccentricity'' $e_{\rm free}$:
\begin{align}
	&e_{\rm forced} = \frac{\nu}{\omega} e_{\rm p} \simeq \frac{5}{4} \alpha e_{\rm p};\\ &e_{\rm free} = |\mathcal{E}_0 - \frac{\nu}{\omega} \mathcal{E}_{\rm p}|. \label{eq:eforced}
\end{align}

\subsection{Ensemble of particles}

If the debris belt were infinitely narrow initially, then it would remain an infinitely narrow belt undergoing coherent precession with an eccentricity oscillating between $|e_{\rm forced} - e_{\rm free}|$ and $(e_{\rm forced} + e_{\rm free})$. However, since a real belt inevitably has a non-zero width, the ensemble of test particles will have a range of precession frequencies $\omega$ linked to their semi-major axis distribution. The belt will lose its coherence in multiple precession periods, and rings of different eccentricities and longitudes of periastron will co-exist. The resulting structure is a belt of mean eccentricity 
\begin{equation}
	\langle e \rangle = e_{\rm forced}, \label{eq:emean}
\end{equation}
and width
\begin{equation}
	\Delta r \simeq 2 \bar{a} e_{\rm free} + \Delta a, \label{eq:da}
\end{equation}
where $\bar{a}$ and $\Delta a$ are the mean value and spread of the semi-major axes of the test particles.

\begin{figure*}
	\centering
	\includegraphics[width=0.45\linewidth]{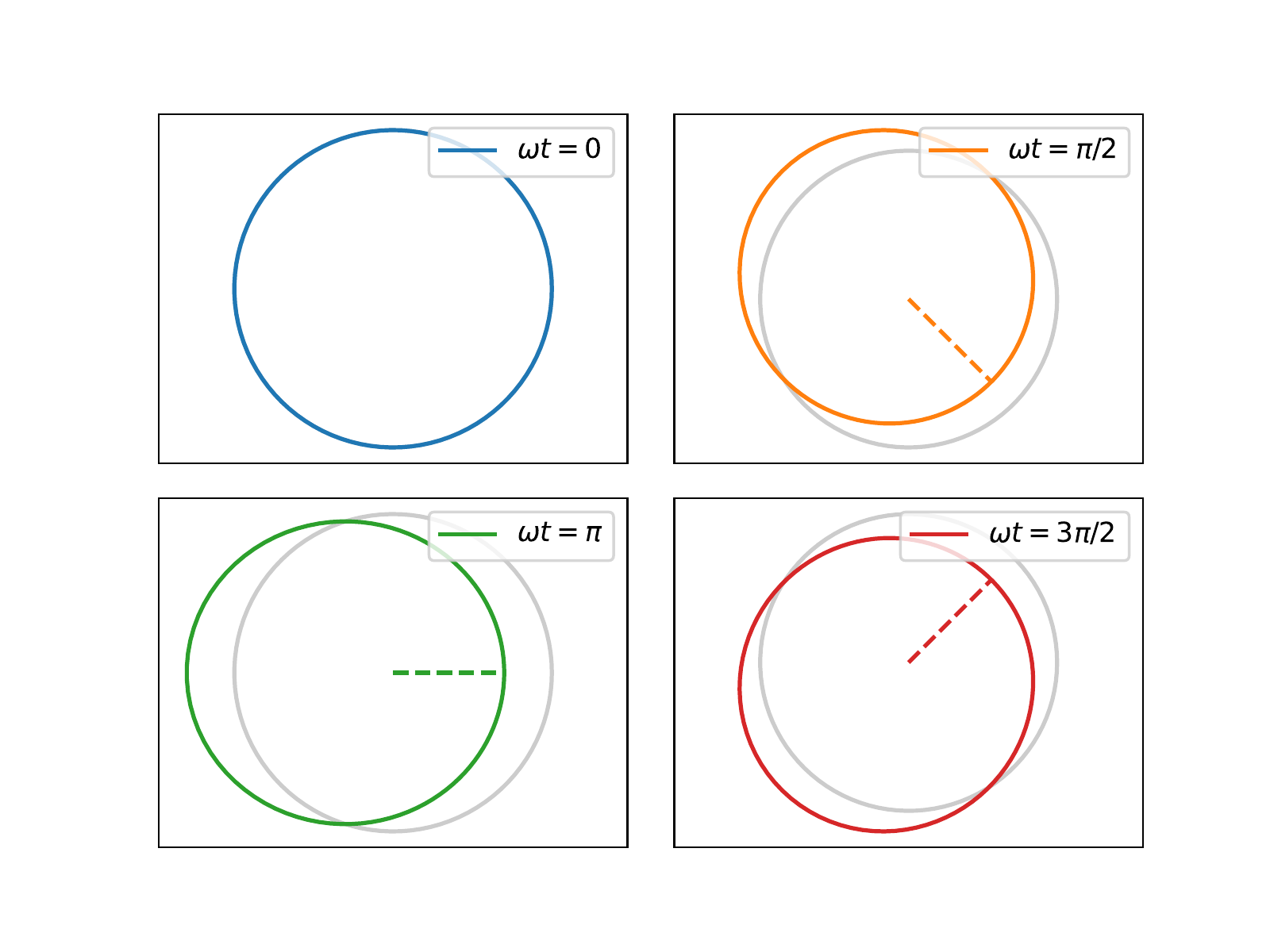}
	\includegraphics[width=0.45\linewidth]{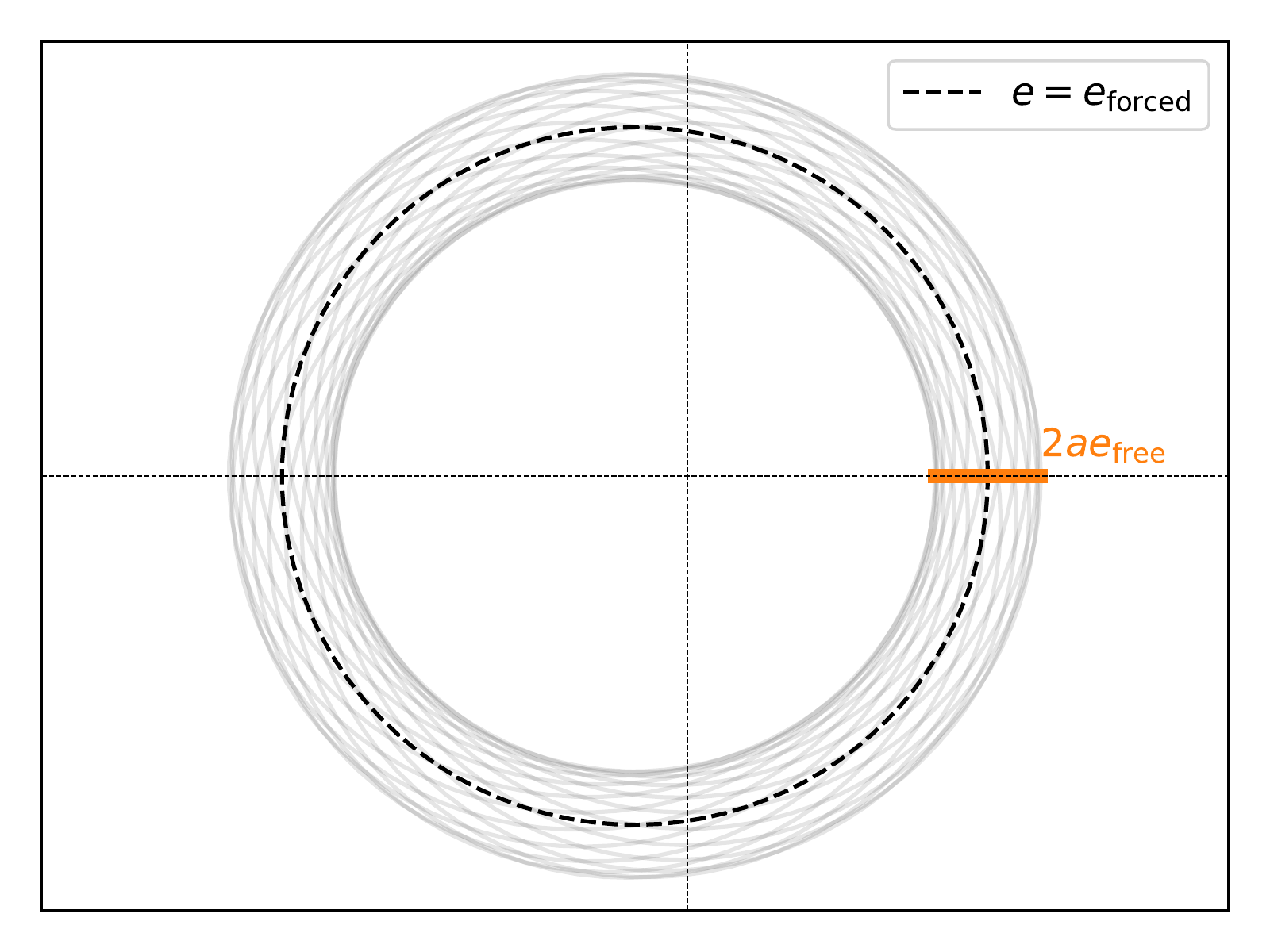}
	\caption{Schematic of the standard picture of an initially circular belt of test particles excited by an inner injected planet of eccentricity $e_{\rm p}$ (equation~\ref{eq:classical}). Left: The coloured orbits correspond to different phases of the precession. The light grey orbit represents the initial circular belt for reference. Right: Superposition of $50$ orbits with incoherent phases, but similar semi-major axes. It can be approximated by an ellipse with eccentricity $e_{\rm forced}$ and width $2 a e_{\rm free}$ (see equations~\ref{eq:rav} and \ref{eq:deltar}).} \label{fig:classical}
\end{figure*}

To show equations~\eqref{eq:emean} and \eqref{eq:da} explicitly, we consider an ensemble of particles with similar semi-major axes $a \simeq \bar{a}$ and write their eccentricities as
\begin{equation}
	\mathcal{E}(\mathcal{W}) \equiv e(\mathcal{W}) \exp\left[i\varpi(\mathcal{W})\right] =  e_{\rm free} \exp(i\mathcal{W}) + e_{\rm forced}.
\end{equation}
At a given time $t$, different particles would have different precession phases $\mathcal{W} = \omega t$ (because of the spread in $a$ and $\omega$). A given orbit characterized by $\mathcal{W}$ has a trajectory described by the polar equation
\begin{align}
	r(\theta, \mathcal{W}) &{}= \frac{\bar{a} \left[1-e^2(\mathcal{W})\right]}{1 + e(\mathcal{W})\cos\left[\theta-\varpi(\mathcal{W})\right]}\nonumber\\
	&{} = \bar{a} \left[ 1 - e_{\rm free} \cos\left(\mathcal{W}-\theta\right) - e_{\rm forced} \cos\theta \right]+ O(e^2),
\end{align}
where $\theta$ is the polar angle and $r$ the radius. At a given $\theta$, the minimum and maximum radii in the ensemble orbits are given by
\begin{align}
	&r_{\rm min} = r(\theta, \mathcal{W} = \theta) \simeq  \bar{a} \left[ 1 - e_{\rm free} - e_{\rm forced} \cos\theta \right]\\
	&r_{\rm max} = r(\theta, \mathcal{W} = \theta+\upi) \simeq \bar{a} \left[ 1 + e_{\rm free} - e_{\rm forced} \cos\theta \right].
\end{align}
The average radius is then:
\begin{align}
	r_{\rm av}(\theta) = \frac{1}{2}\left(r_{\rm min} + r_{\rm max} \right) = \bar{a}\left(1 - e_{\rm forced} \cos\theta\right). \label{eq:rav}
\end{align}
Thus, the apparent orbit of this ensemble has an eccentricity $e_{\rm forced}$. The width of the ensemble is
\begin{align}
	\Delta r(\theta) = r_{\rm max} - r_{\rm min} = 2 \bar{a} e_{\rm free}.\label{eq:deltar}
\end{align}
These results are illustrated in Figure~\ref{fig:classical}.

\subsection{Eccentricity damping}
\label{sec:eccentricitydamping}

A common hypothesis to explain the narrow eccentric debris belts in extrasolar systems (see Section~\ref{sec:introduction}) is the presence of an eccentricity-damping force (e.g. due to the friction from the surrounding gas) acting on the test particles. Equation~\eqref{eq:eqdiff} becomes
\begin{equation}
	\dv{\mathcal{E}}{t} = \left(i\omega - \frac{1}{\tau_{\rm d}}\right)\mathcal{E} - i \nu \mathcal{E}_{\rm p}, \label{eq:eqdiff_damping}
\end{equation}
where $\tau_{\rm d}$ is the eccentricity damping time. For $\mathcal{E}_0 = 0$, equation~\eqref{eq:eqdiff_damping} has the solution
\begin{equation}
	\mathcal{E}(t) = \frac{\nu}{\omega} \mathcal{E}_{\rm p} \frac{\tau_{\rm d}\omega}{i+\tau_{\rm d}\omega} \left[1-\exp \left(i\omega t - \frac{t}{\tau_{\rm d}} \right)\right]. \label{eq:damping}
\end{equation}
For $t \gg \tau_{\rm d}$, $\mathcal{E}(t)$ approaches a fixed value with zero free eccentricity:
\begin{equation}
	\mathcal{E}(t) \xrightarrow[t\to\infty]{} \frac{\nu}{\omega}\mathcal{E}_{\rm p} \frac{\tau_{\rm d}\omega}{i+\tau_{\rm d}\omega}. \label{eq:shift}
\end{equation}
Not that for $\omega\tau_{\rm d} \gg 1$, this asymptotic eccentricity reduces to $e_{\rm forced} = (\nu/\omega) e_{\rm p}$. But if $\omega\tau_{\rm d}$ is not much larger than unity, $\mathcal{E}(t\to \infty)$ is not equal to $e_{\rm forced}$ and the eccentric disc is not apsidally aligned with the planetary perturber.

\subsection{Planetary Mass Growth}

Now consider the case where the planet has a constant eccentricity, but its mass $m_{\rm p}(t)$ is growing with time.
Both frequencies $\omega$ and $\nu$ then depend on time. However, since the ratio $\nu(t)/\omega(t)$ is constant, equation~\eqref{eq:solution} can be integrated out explicitly, which gives
\begin{equation}
	\mathcal{E}(t) = \frac{\nu}{\omega} \mathcal{E}_{\rm p} + \left(\mathcal{E}_0 - \frac{\nu}{\omega} \mathcal{E}_{\rm p} \right) \exp\left[i \mathcal{W}(t)\right].\label{eq:massgrowth}
\end{equation}

Comparing to equation~\eqref{eq:classical}, we see that although the time evolution is different from the constant-$m_{\rm p}$ case, the eccentricity has exactly the same trajectory in the complex plane. This is a fundamental property of secular perturbations: their amplitude depends on the semi-major axis ratio and planet eccentricity but does not depend on the planet mass (which only impacts the timescale). Therefore, mass growth alone cannot reduce the free eccentricity and the width of the debris belt.

\subsection{Planetary Eccentricity Growth}
\label{sec:eccentricitygrowth}

\begin{figure}
	\centering
	\includegraphics[width=\linewidth]{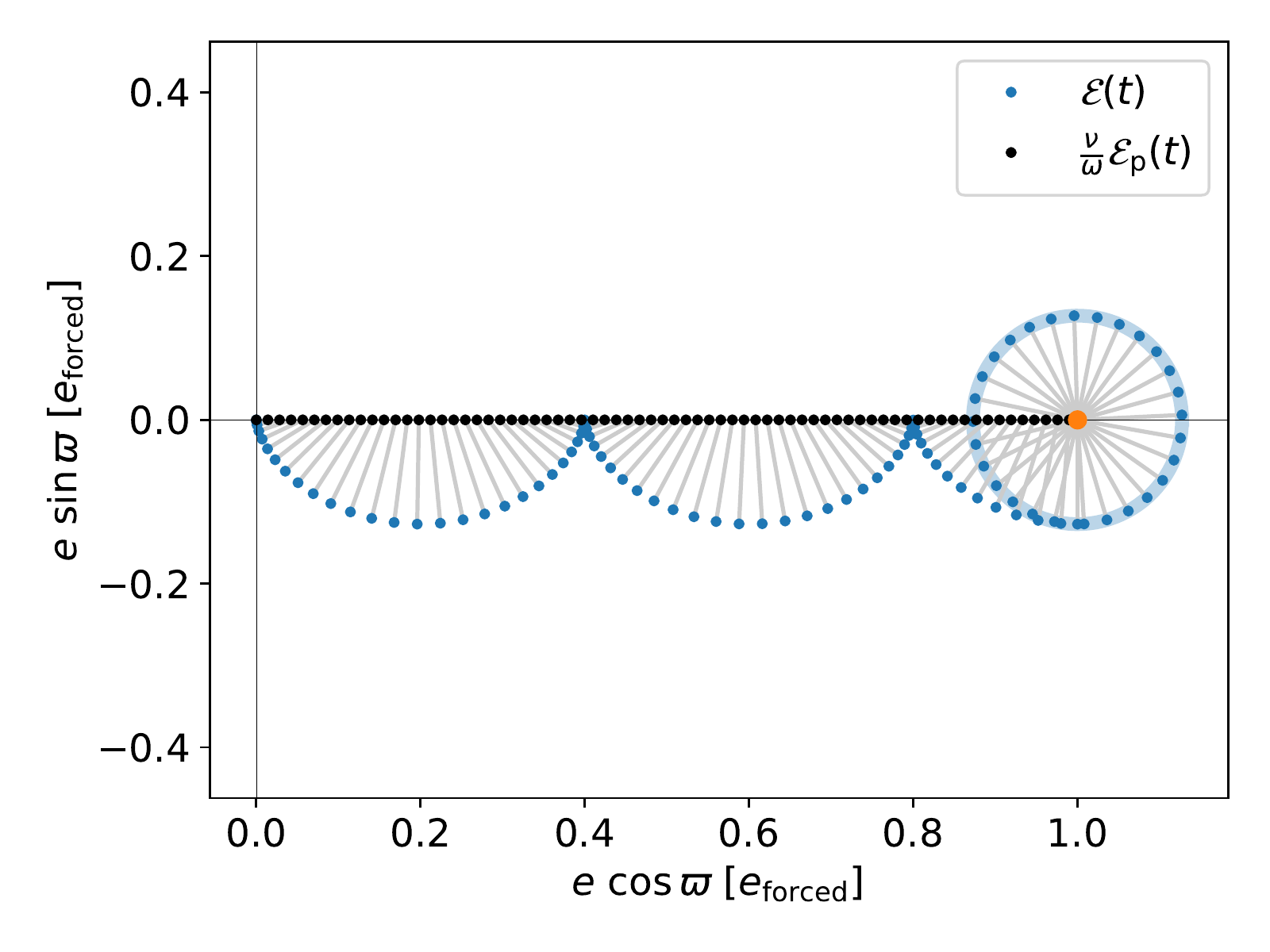}
	\caption{Evolution of the complex eccentricity of a test particle (blue dots, equations~\ref{eq:epstau}) exterior to a planet with a linear eccentricity growth (equations~\ref{eq:ecc}), in units of the final forced eccentricity. We adopt the growth time $t_{\rm p} = 2.5 ~t_\omega$, where $t_\omega = 2\upi/\omega$. The test particle eccentricity is initially zero, its evolution is represented by $100$ dots sampled uniformly between $t = 0$ and $t_{\rm p}+t_\omega$. The corresponding forced eccentricity $\frac{\nu}{\omega}\mathcal{E}_{\rm p}(t)$ is represented by black dots linked with a grey line. The orange dot shows the final forced eccentricity (equation~\ref{eq:eforced_egrowth}), and the radius of the light blue circle is the final free eccentricity (equation~\ref{eq:efree_egrowth}). At each time-step, the complex eccentricity of the test particle (in blue) rotates anticlockwise around the current forced eccentricity (in black), which results in the half ellipse. At $x=0.4$, $e_\mathrm{forced}$ `overtakes' $\mathcal{E}$, so the half ellipse starts the cycle anew.}\label{fig:ecomplex}
\end{figure}

Now consider the case where the planet eccentricity grows linearly in time until reaching the final value $e_{\rm p, f}$, i.e.
\begin{align}
	e_{\rm p}(t) = \begin{cases}
		e_{\rm p, f} \frac{t}{t_{\rm p}} &\text{for } 0 \leq t \leq t_{\rm p}\\
		 e_{\rm p, f} &\text{for } t > t_{\rm p}.
	\end{cases}\label{eq:ecc}
\end{align}
We assume $\varpi_{\rm p} = 0$ for simplicity. Again, we can integrate equation~\eqref{eq:solution} to obtain the complex eccentricity of the test particle:
\begin{align}
	&\mathcal{E}(t) = \nonumber\\&\begin{cases}
		e_\mathrm{forced} \frac{1}{t_{\rm p}} \left[t +\frac{i}{\omega}(\exp i\omega t -1)  \right]  &\text{for } 0 \leq t \leq t_{\rm p} \\
		e_\mathrm{forced} +  \left[\mathcal{E}(t_{\rm p}) - e_{\rm forced}\right] \exp\left[i\omega (t-t_{\rm p})\right] &\text{for } t > t_{\rm p},
	\end{cases}  \label{eq:epstau}
\end{align}
where
\begin{equation}
	e_\mathrm{forced} = \frac{\nu}{\omega} e_{\rm p, f} \label{eq:eforced_egrowth}
\end{equation}
is the ``final'' forced eccentricity. Clearly, the free eccentricity of the test mass at $t>t_{\rm p}$ is
\begin{align}
	e_{\rm free} ={}& \left|\mathcal{E}(t_{\rm p}) - e_{\rm forced}\right|\nonumber\\
	={}& e_{\rm forced} \left|\frac{2\sin\left(\frac{\omega t_{\rm p}}{2}\right)}{\omega t_{\rm p}}\right|\label{eq:efree_egrowth}
\end{align}
From equation~\eqref{eq:efree_egrowth}, we see that the free eccentricity can be lower than the forced eccentricity if $\omega t_{\rm p} \gtrsim 1$, i.e. if the planet's eccentricity growth is sufficiently slow. An example of the test particle eccentricity evolution in the complex plane in such a case is shown in Figure~\ref{fig:ecomplex}.

\begin{figure}
	\centering
	\includegraphics[width=\linewidth]{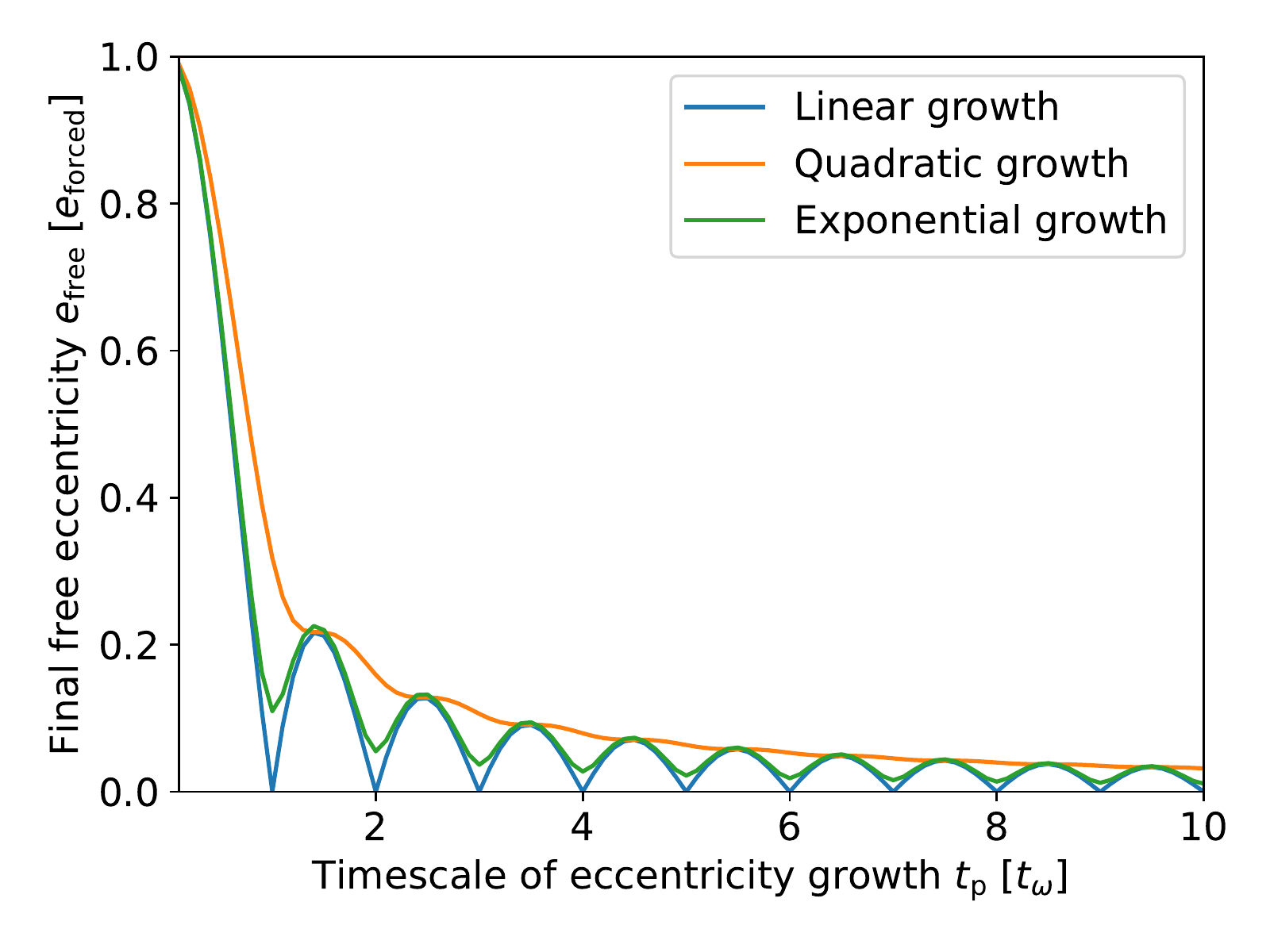}
	\caption{Final free eccentricity of an exterior test particle (in units of the final forced eccentricity, equation~\ref{eq:eforced_egrowth}) as a function of the planet eccentricity time growth $t_{\rm p}$ (in units of the precession period $t_\omega = 2\upi/\omega$, equation~\ref{eq:efree_egrowth}), for three different types of eccentricity growth functions (linear, quadratic, exponential). In all three cases, the envelope of the final free eccentricity decreases as $1/t_{\rm p}$. The lower the free eccentricity, the narrower the debris belt (equation~\ref{eq:da}).}\label{fig:efree}
\end{figure}

A similar analysis can be conducted for different eccentricity functions: quadratic $\left[e_{\rm p} \propto (t/t_{\rm p})^2\right]$ and exponential $\left[e_{\rm p} \propto \exp(t/t_{\rm p})\right]$. The results are shown in Figure~\ref{fig:efree}. We note that the final free eccentricity depends less on the growth functional form than on the timescale $t_{\rm p}$: if $t_{\rm p}$ is greater than the precession time $t_\omega = 2\upi/\omega$, then the free eccentricity is less than half of the final forced eccentricity, and so is the expected debris belt width.

\section{Interaction with protoplanetary disc}
\label{sec:ppdisc}

In the previous section, we have presented a general analysis for the free eccentricity of a test particle under the influence of an evolving planet. This suggests two mechanisms of reducing the width of eccentric debris discs: eccentricity damping forces acting on the planetesimals and slow eccentricity growth of the perturbing planet. In this section, we apply these general mathematical solutions to the physical picture of ``planet+disc'' evolution, in which the gaseous protoplanetary disc damps the eccentricity of the planetesimals while exciting the eccentricity of the giant planet.

\subsection{Planet eccentricity growth in gas disc}
\label{sec:planetembedded}

Massive (gap-opening) planets interacting with protoplanetary discs can experience eccentricity growth \citep[e.g.;][]{goldreichEccentricityEvolutionPlanets2003}. Recent numerical simulations indicate that the growth time can be as long as $10^5$--$10^6$ planet orbits \citep[e.g.;][]{ragusaEccentricityEvolutionPlanetdisc2018}. On the other hand, the precession time of the planetesimals driven by the planet is (see equation~\ref{eq:frequency}):
\begin{equation}
	t_\omega = \frac{2\upi}{\omega} \simeq 6\times 10^4~P_{\rm p} \left(\frac{a}{3a_{\rm p}}\right)^\frac{7}{2} \left(\frac{\mu_{\rm p}}{10^{-3}}\right)^{-1}, \label{eq:tw}
\end{equation}
where $P_{\rm p}$ is the orbital period of the planet.

We have seen in Figure~\ref{fig:efree} that as long as $t_{\rm p} \gtrsim t_\omega$, the free eccentricity of the test particle is less than $20\%$ the forced eccentricity, which would significantly reduce the physical width of the planetesimal belt. Equation~\eqref{eq:tw} indicates that to have $t_\omega \lesssim 10^5~P_{\rm p}$ would require $a/a_{\rm p} \lesssim 3.4~(\mu_{\rm p}/10^{-3})^{2/7}$, i.e. the perturbing planet must be very close to the planetesimal belt.

Since the planetary eccentricity growth time must be less than the lifetime $t_{\rm life}$ of the disc ($\sim$ Myrs), a necessary condition to reduce the width of debris belt is $t_{\omega} < t_{\rm life}$. This gives
\begin{equation}
	\left(\frac{P_{\rm p}}{10~\mathrm{yr}}\right) \left(\frac{a}{3a_{\rm p}}\right)^\frac{7}{2} \lesssim \left(\frac{\mu_{\rm p}}{10^{-3}}\right).
\end{equation}
Therefore, this process requires the planet to be relatively close to both the star and the planetesimal belt. 

\subsection{Planetesimal eccentricity damping in gas disc}
\label{sec:tpembedded}

Let us now consider the situation where the debris belt of planetesimals is embedded in a gaseous disc, while the planetary orbit remains constant. Such a configuration could occur in transition discs, where the planet lies inside a gas-free cavity, like has been observed in the benchmark system PDS 70 \citep{mullerOrbitalAtmosphericCharacterization2018}. 

In Section~\ref{sec:eccentricitydamping}, we considered the effect of linear eccentricity damping ($\dot{e} = - e/\tau_{\rm d}$) on the free eccentricity of test particles. Such linear damping does not always apply to planetesimal-gas disc interaction. In fact, the frictional force acting on a planetesimal of radius $R$ from a gas disc of density $\rho_{\rm g}$ takes the form \citep[e.g., ][]{grishinApplicationGasDynamical2016}:
\begin{equation}
	\vec{F} = -\frac{1}{2} C_{\rm d} \upi R^2 \rho_{\rm g} |\Delta v| \vec{\Delta v}, \label{eq:frictionforce}
\end{equation}
where $\vec{\Delta v}$ is the difference between the planetesimal velocity and the Keplerian velocity of the surrounding gas, and $C_{\rm d}$ is the drag coefficient ($\sim 1$ for planetesimals of $R \sim 1$--$100$ km). 
The gas density $\rho_{\rm g}$ is related to the surface density $\Sigma_{\rm g}$ and disc scale height $H$ by 
\begin{equation}
	\rho_{\rm g} = \frac{\Sigma_{\rm g}}{2 H}.
\end{equation}
Using the disc aspect ratio $h = H/a = 0.1$ and assuming $\Sigma_{\rm g} = 2.10^{3}~\mathrm{g.cm^{-2}} (a/1~\mathrm{au})^{-3/2}$ \citep[the minimum mass solar nebula;][]{hayashiStructureSolarNebula1981}, we get
\begin{equation}
	\rho_{\rm g} \simeq 10^{-9} ~\mathrm{g.cm^{-3}} \left(\frac{a}{1~\mathrm{au}}\right)^{-\frac{5}{2}}.
\end{equation}
The frictional force induces an eccentricity damping (see Appendix~\ref{sec:appendix_eccentricitydamping})
\begin{equation}
	\frac{\dot{e}}{e} = -\frac{1}{\tau_{\rm d}(e)} = -\frac{e}{\tau_0}, \label{eq:nonlinearfriction}
\end{equation}
with
\begin{align}
	\tau_{0} &{}\simeq \frac{0.5}{C_{\rm d}} \frac{\rho_{\rm pl}}{\rho_{\rm g}} \frac{R}{a} ~P\\
	&{}\simeq 10^8~\mathrm{yr}\left(\frac{R}{10~\mathrm{km}}\right) \left(\frac{a}{100~\mathrm{au}}\right)^3 \left(\frac{M_*}{1~\mathrm{M_\odot}}\right)^{-\frac{1}{2}} \nonumber\\
	&\qquad \times C_{\rm d}^{-1}\left(\frac{\rho_{\rm pl}}{2~\mathrm{g.cm^{-3}}}\right)  \left(\frac{\rho_{\rm g,0}}{10^{-9}~\mathrm{g.cm^{-3} }}\right),\label{eq:tau0}
\end{align}
where $P$ is the orbital period, $\rho_{\rm pl}$ the bulk density of the planetesimal and $\rho_{\rm g, 0}$ the value of the gas density $\rho_{\rm g}$ at $1$ au. Since all quantities have a wide range of possible values, the damping constant $\tau_0$ is uncertain, and can range from $10^2$ to $10^{10}$ years (depending in particular on the location of the planetesimal belt).

\begin{figure}
	\centering
	\includegraphics[width=\linewidth]{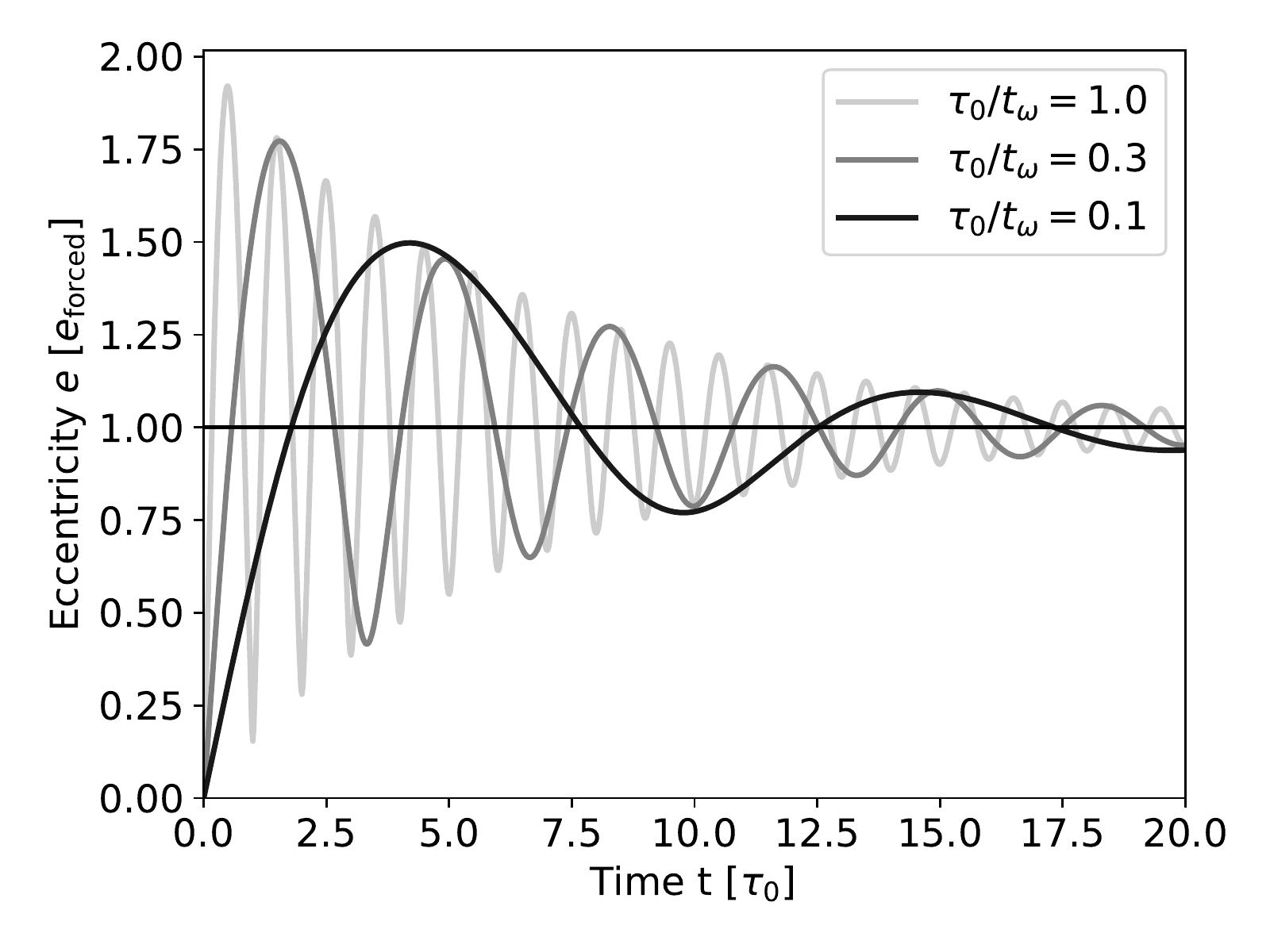}
	\caption{Eccentricity evolution of a test particle exterior to a planet companion, with the particle experiencing a non-linear friction force with the eccentricity damping timescale $\tau_{\rm d}(e) = \tau_{0}/e$ (equation~\ref{eq:nonlinearfriction}). Each line corresponds to a different ratio between $\tau_{0}$ and the precession timescale $t_\omega = 2\upi/\omega$. All cases are for $e_{\rm forced} = 0.1$. We see that the damping acts on the timescale of about $\tau_{\rm d}(e_{\rm forced}) \simeq 10 ~\tau_{0}$. }\label{fig:efree_friction}
\end{figure}

The evolution of the complex eccentricity $\mathcal{E}$ of a planetesimal driven by the perturbing planet is given by
\begin{equation}
	\dv{\mathcal{E}}{t} = i\omega\mathcal{E} - \frac{|\mathcal{E}|}{\tau_{\rm 0}} \mathcal{E} - i \nu \mathcal{E}_{\rm p}.
\end{equation}
This differential equation is non-linear, and cannot be solved analytically. We use the \textsc{Python SciPy} library to solve it numerically, and plot the results of three different eccentricity evolutions in Figure~\ref{fig:efree_friction}.
We see that the test particle's eccentricity is driven toward an equilibrium value close to $e_{\rm forced} = (\nu/\omega) e_{\rm p}$ (see equation~\ref{eq:eeq} in Appendix~\ref{sec:appendix_eccentricitydamping} for the exact value) on the timescale of $\tau_0/e_{\rm forced}$. When the protoplanetary disc dissipates, the free eccentricity of the planetesimals will be roughly equal to the difference between this equilibrium and $e_{\rm forced}$: the width of the belt will thus be greatly reduced compared to the standard picture. However, this process requires the protoplanetary disc to live long enough for the damping to act, that is $\tau_{0}/e_{\rm forced} \lesssim t_{\rm life}$. From equation~\eqref{eq:tau0}, we see that this requires the debris belt to be close-in ($a \lesssim 10$ au).

\begin{figure}
	\centering
	\includegraphics[width=\linewidth]{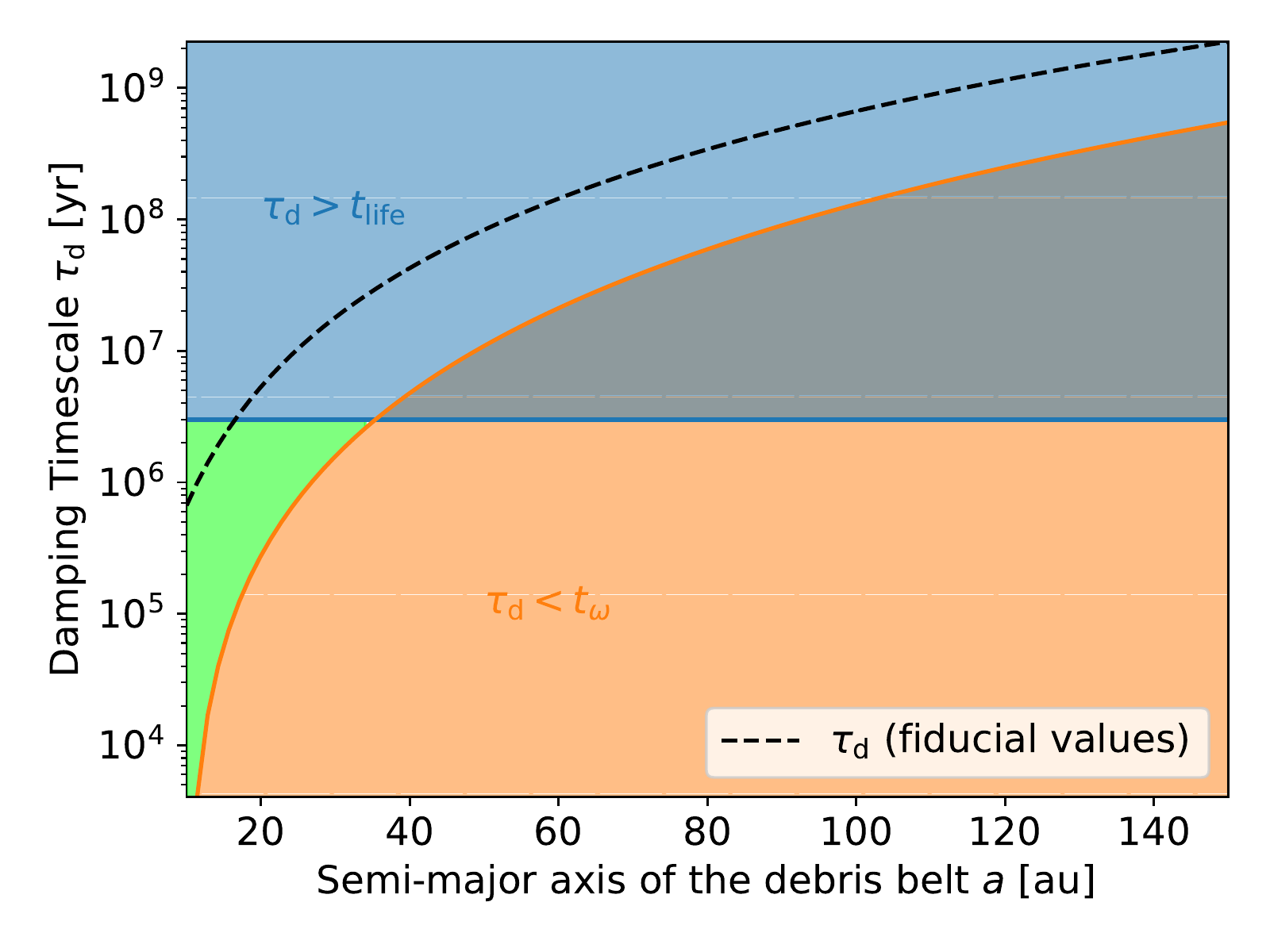}
	\caption{Parameter space ($a$, $\tau_{\rm d}$) that successfully decreases the debris belt free eccentricity (in green). The blue zone corresponds to an inefficient damping, which would require more time than the lifetime of the protoplanetary disc. The orange zone corresponds to a strong damping ($\tau_{\rm d} < t_\omega$, with $\mu_{\rm p} = 10^{-3}$ and $a_{\rm p} = 10$ au), which would shift the forced eccentricity away from its gas-free value. The black dashed line corresponds to $\tau_{\rm d} = \tau_{0}/e_{\rm forced}$ following equation~\eqref{eq:tau0} with the fiducial parameters and $e_{\rm forced} = 0.1$. For the parameters we chose, the free eccentricity can be reduced only for close-in debris belts ($a \lesssim 10$ au). In order to increase the efficiency of this mechanism to wider belts, we would need a more efficient damping (i.e. smaller planetesimals or higher gas density) and a smaller precession period (i.e. planet closer to the belt or with a higher mass).}\label{fig:tau_vs_a}
\end{figure}

Moreover, a strong damping force can shift the equilibrium eccentricity away from the forced eccentricity (as described in Section~\ref{sec:eccentricitydamping}), and subsequently hinder the reduction of the free eccentricity. Although the non-linear aspect of the damping (dependence of $\tau_{\rm d}$ on the eccentricity) reduces this effect, it becomes significant for $\tau_{0}\omega \lesssim e_{\rm forced}$ (see Appendix~\ref{sec:appendix_eccentricitydamping}). For the fiducial values we consider in this paper (equations~\ref{eq:tw} and \ref{eq:tau0} with $a = 10$--$100$ au), the damping timescale is larger than the precession timescale, so that the difference between equilibrium and forced eccentricities is negligible. Both timescale constraints are represented on Figure~\ref{fig:tau_vs_a}.

\section{Planet-Planet Scattering}
\label{sec:ppscattering}

In the previous section, we showed that the interaction between the planet and debris belt and their surrounding protoplanetary disc can reduce the planet-induced free eccentricity of the planetesimals and thus increase the coherence of the belt. However, this requires either a strong and sustained frictional force or a protoplanetary disc lifetime a few times larger than the precession period of the planetesimals. These conditions may not be met for wide debris belts ($\gtrsim 100$ au) including several observed systems.

In this section, we examine another process to decrease the free eccentricity of the planetesimals, by considering the scenario in which the perturbing planet gains its eccentricity through planet-planet scattering, after the dispersal of the protoplanetary disc. Strong gravitational scatterings is a leading mechanism to produce extrasolar giant planets on eccentric orbits \citep[e.g.;][]{chatterjeeDynamicalOutcomesPlanetPlanet2008,juricDynamicalOriginExtrasolar2008,fordOriginsEccentricExtrasolar2008,andersonSituScatteringWarm2020,liGiantPlanetScatterings2021}. In this scenario, the eccentricities of the planets change in an irregular way (approximatively following random walks) until one of the planets is ejected. Our goal is to understand how the eccentricity of a planetesimal evolves during the planet-planet scattering process.

\subsection{Numerical random-walk model}
\label{sec:randomgrowth_numerical}

$N$-body simulations suggest that planet-planet scatterings can be modelled as a random process \citep{puStrongScatteringsCold2021}. Here we present a toy model where the planet eccentricity grows following a discrete random walk. We suppose that the planet has an initially circular orbit, and that we know its final eccentricity $e_{\rm p,f}$ at the end of the scattering process (e.g. $e_{\rm p, f}$ is the observed eccentricity of the perturbing planet around a debris disc).

The walk thus consists of a collection of instantaneous kicks in the 2D complex plane, beginning at $0$ and ending at $\mathcal{E}_{\rm p} = e_{\rm p,f}$ (fixed) after $N_{\rm p}$ steps. The magnitude of each kick is sampled from a Gaussian distribution of scale $\sqrt{2}\Delta e_{\rm p}$\footnote{ Note that $e_{\rm p}\cos\varpi_{\rm p}$ and  $e_{\rm p}\sin\varpi_{\rm p}$ each are sampled from a Gaussian distribution of scale $\Delta e_{\rm p}$, so that $\Delta e_{\rm p}$ is the equivalent to $\sigma_{\rm p}$ in Section 4.2.} and the direction of the kick is random. The scale of the kick and the final value of the planet eccentricity give a most likely value for the number of kicks before the end of the scattering
\begin{equation}
	\bar{N}_{\rm p} \equiv \frac{e_{\rm p,f}^2}{2 \Delta e_{\rm p}^2}.
\end{equation}
The actual number of kicks $N_{\rm p}$ has a distribution around $\bar{N}_{\rm p}$.

The complex eccentricity of the test particle (the planetesimal) will try to follow the evolution of the planet eccentricity. The result depends on the number of kicks that occur within one planetesimal precession period, defined as $N_\omega$ [if the typical scattering step lasts $\Delta t$, then $N_\omega \approx (\omega\Delta t)^{-1}$ ]. If $N_\omega$ is much less than the total number of steps $N_{\rm p}$, then the whole scattering process can be seen as adiabatic. On the other hand, if $N_\omega$ is much larger than the total number of steps $N_{\rm p}$, then the process can be considered instantaneous.

According to \cite{puStrongScatteringsCold2021}, the number of close encounters $N_{\rm p}$ of a scattering process between two giant planets follows a Lévy distribution peaking at $\bar{N}_{\rm p}$. In our study, we fix $\bar{N}_{\rm p} = 10^3$ \citep[the value of $\bar{N}_{\rm p}$ depends on the masses of the two planets, see Figs.~3-4 of][]{puStrongScatteringsCold2021}, and we thus produce a set of random eccentricity kicks following such a distribution (see Appendix~\ref{sec:randomwalk_method} for details).

We then use $N_\omega$ to compute the evolution of the complex eccentricity $\mathcal{E}$ of the test particle using equation~\eqref{eq:solution} (assuming $\mathcal{E}_{\rm p}$ remains constant between kicks), and use its final value $\mathcal{E}(N_{\rm p})$ to derive the free eccentricity $e_{\rm free}$ as a function of the final forced eccentricity $e_{\rm forced}$:
\begin{align}
	&e_{\rm forced} = \frac{\nu}{\omega} \mathcal{E}_{\rm p} (N_{\rm p}) = \frac{\nu}{\omega} e_{\rm p,f} \label{eq:eforced_random},\\
	&e_{\rm free} = |\mathcal{E}(N_{\rm p}) - e_{\rm forced}|. \label{eq:efree_random}
\end{align}
Note that the forced eccentricity does not depend on the evolutionary path taken by the planet eccentricity during the scattering process. However, the free eccentricity does, and depends in particular on $N_{\rm p}$ and $N_\omega$.

\begin{figure}
	\centering
	\includegraphics[width=\linewidth]{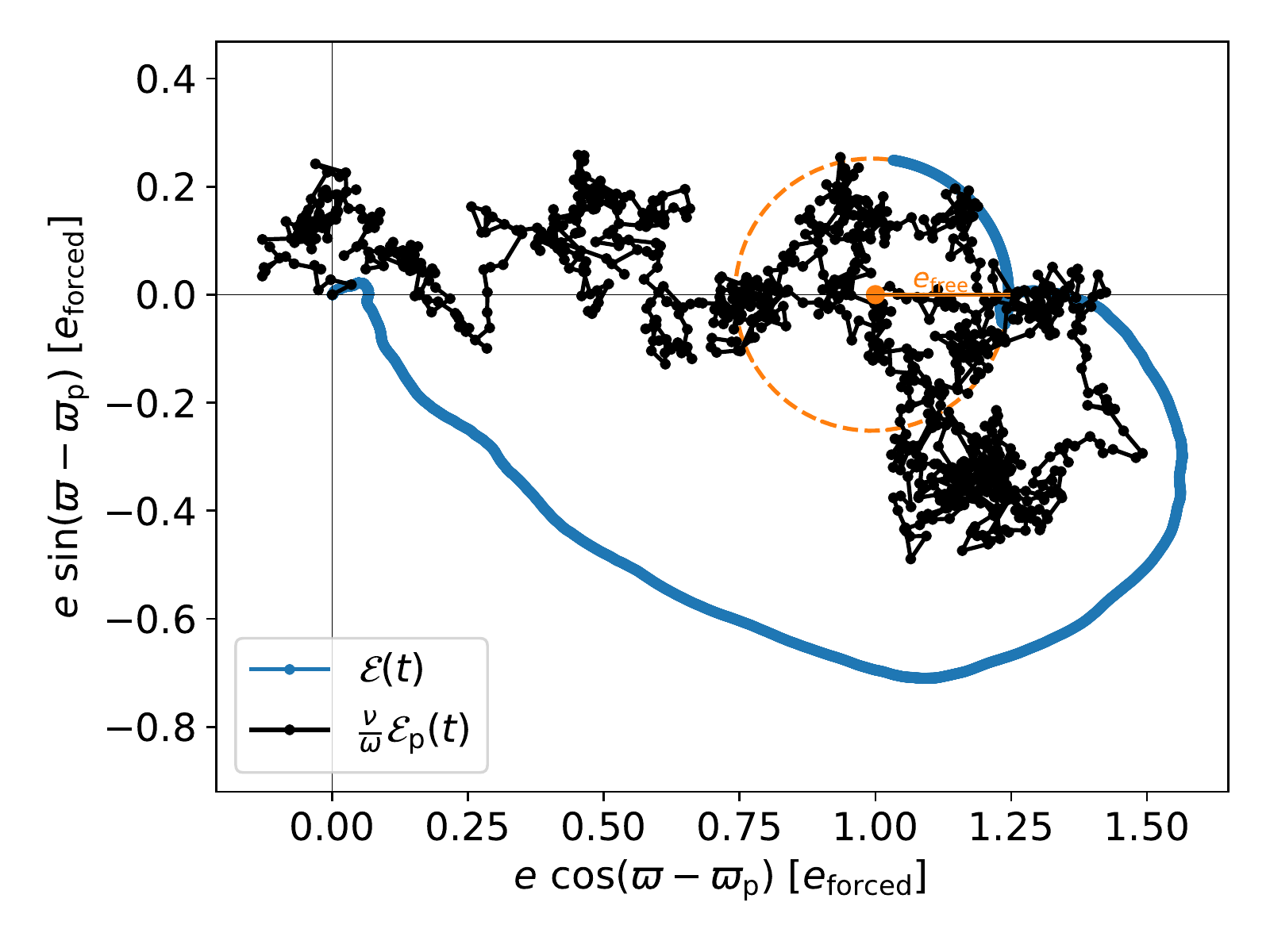}\\
	\includegraphics[width=\linewidth]{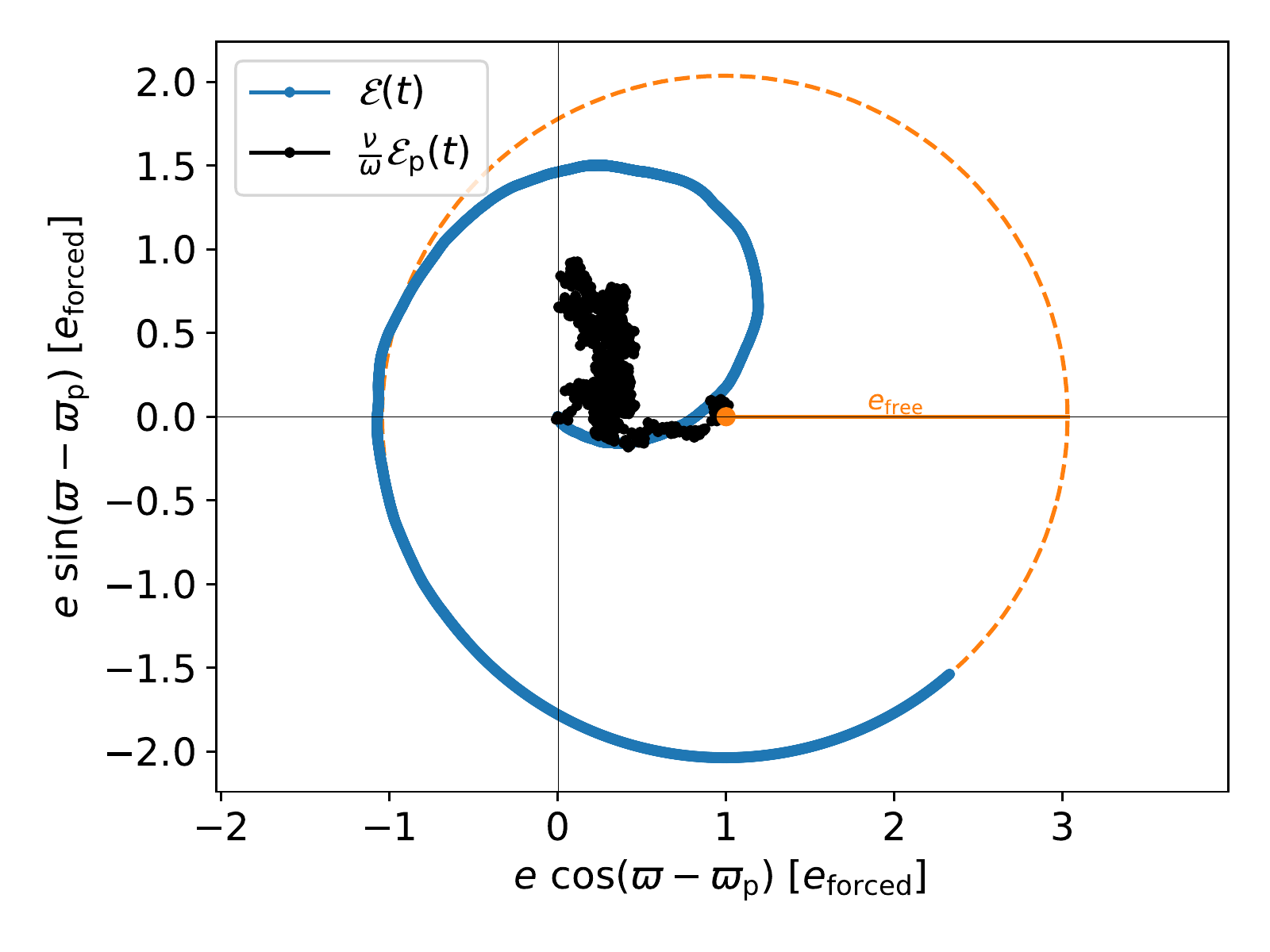}
	\caption{Examples of the evolution of the complex eccentricity of the perturbing planet (in black) which undergoes planet-planet scattering, and the corresponding evolution of an outer test particle (in blue). These evolutions are computed using the numerical random-walk model presented in Section~\ref{sec:randomgrowth_numerical} with $N_\omega = \bar{N}_{\rm p} = N_{\rm p} =  1000$. The orange dotted line represents the evolution of the test particle at the end of the scattering process. The example in the upper panel leads to a particle having a small free eccentricity, the example on the bottom panel leads to a high free eccentricity.}\label{fig:ecomplex_randomgrowth}
\end{figure}

Two examples of random walks are shown in Figure~\ref{fig:ecomplex_randomgrowth}, one leading to a free eccentricity of the test particle less than the forced eccentricity, the other leading to a larger free eccentricity. Note that in this stochastic model, the free eccentricity can be either smaller or larger than $e_{\rm forced}$, contrary to the standard picture (see Section~\ref{sec:classical}). 

\begin{figure}
	\centering
	\includegraphics[width=\linewidth]{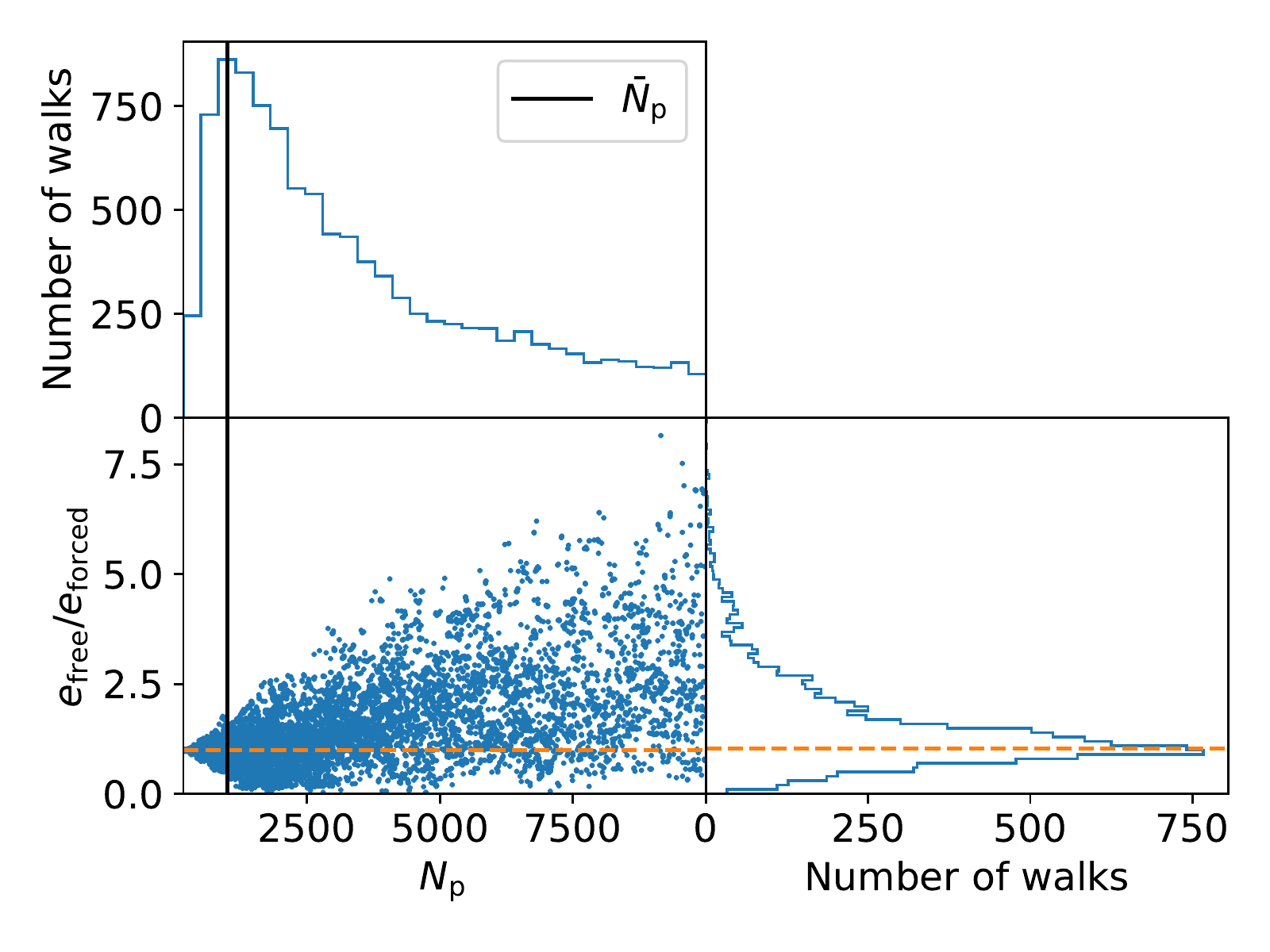}
	\caption{Final free eccentricity of the planetesimal as a function of the number of kicks ($N_{\rm p}$) in the planet-planet scattering process, assuming $N_\omega = 5000$ and $\bar{N}_{\rm p} = 1000$. The dots in the bottom left panel represents the individual results of 10,000 walks, the top left and bottom right panels show the histograms depicting the distributions of $N_{\rm p}$ and the free eccentricities, respectively.}\label{fig:efree_randomgrowthvssteps}
\end{figure}

\begin{figure}
	\centering
	\includegraphics[width=\linewidth]{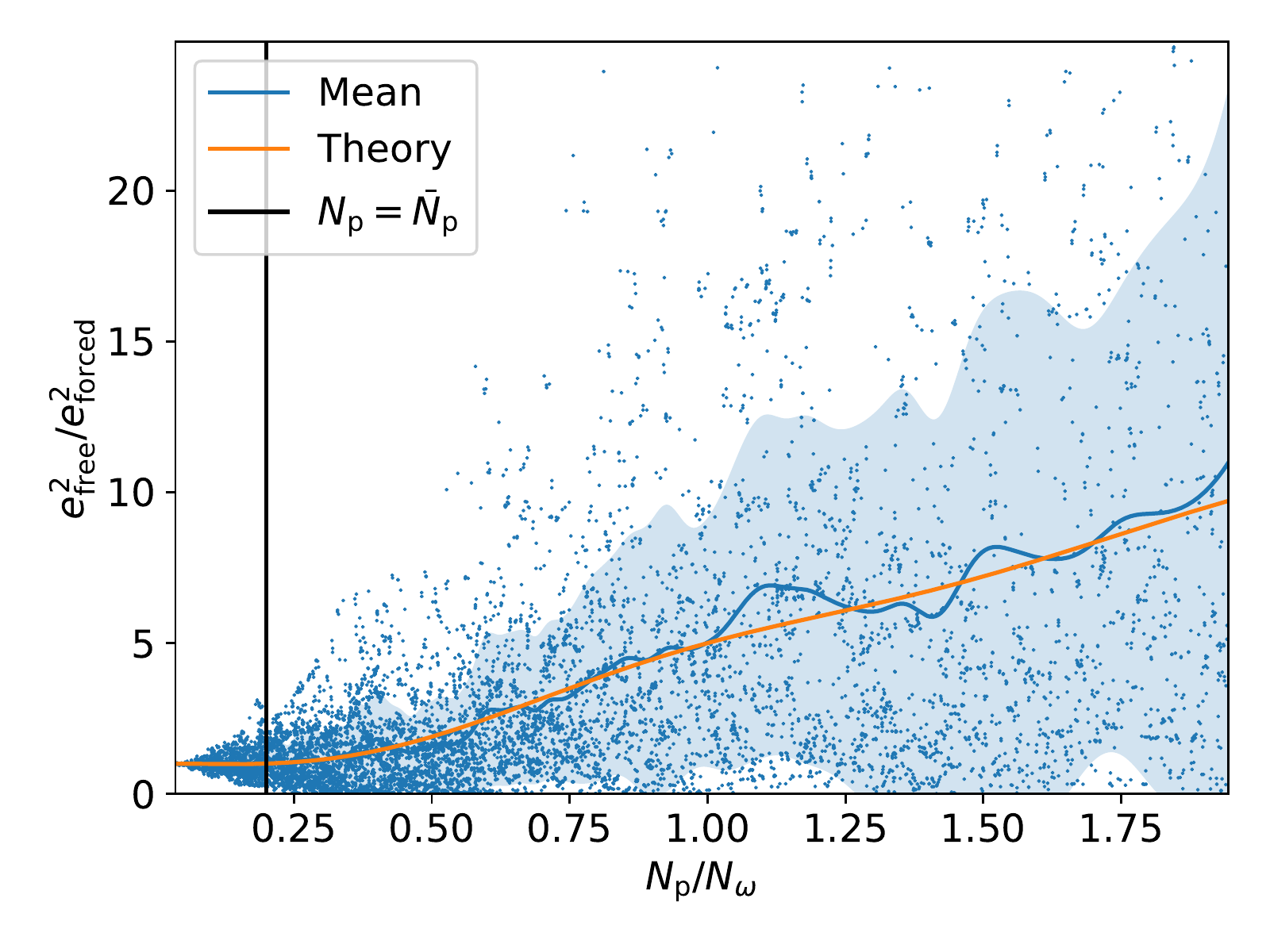}
	\includegraphics[width=\linewidth]{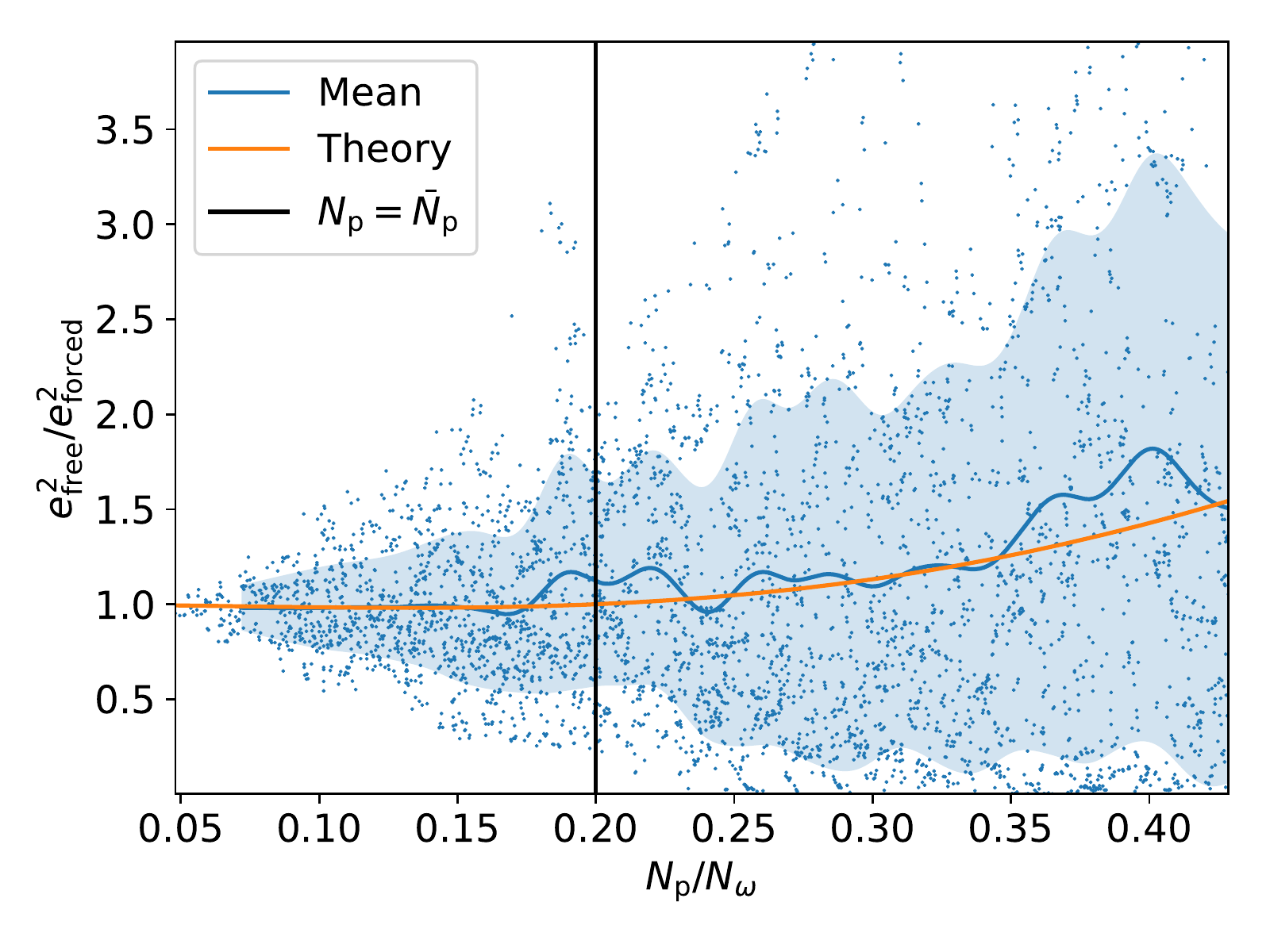}
	\caption{Same as Figure~\ref{fig:efree_randomgrowthvssteps}, except the $x$-axis is the number of kicks normalized by $N_\omega$ and the $y$-axis represents the squared free eccentricity. The blue line and shade correspond to the moving squared average and standard deviation respectively. The theoretical expectation (equation~\ref{eq:efreerms}) is represented in orange. The bottom panel zooms in the zone of low free eccentricity in the upper panel.}\label{fig:efree_randomgrowthvsstepstheory}
\end{figure}

Figure~\ref{fig:efree_randomgrowthvssteps} shows the free eccentricity of the test particle in $10^4$ random walk samples as a function of their total number of steps $N_{\rm p}$. The distribution of the free eccentricities peaks at the forced eccentricity. For sufficiently large $N_{\rm p}$, the average and spread of free eccentricities appear to grow with $N_{\rm p}$. Conversely, if $N_{\rm p}$ is small, then the free eccentricities converge to the $e_{\rm forced}$---This recovers the result of the standard picture discussed in Section~\ref{sec:classical}, where the planet's eccentricity growth is instantaneous.  
In the next subsection, we will demonstrate that the most relevant dependency is with $N_{\rm p}/N_\omega$, and that the root mean square of the free eccentricity has an analytical expression. We thus plot the data set of Figure \ref{fig:efree_randomgrowthvssteps} in the $(N_{\rm p}/N_\omega, e_{\rm free}/e_{\rm forced})$ space in Figure~\ref{fig:efree_randomgrowthvsstepstheory}. For $N_{\rm p} \lesssim \bar{N}_{\rm p}$, we have $\langle e_{\rm free}^2 \rangle \sim e_{\rm forced}^2$, while for $N_{\rm p} \gtrsim \bar{N}_{\rm p}$, $\langle e_{\rm free}^2 \rangle /e_{\rm forced}^2$ is proportional to $N_{\rm p}/N_\omega$. For a fixed $N_{\rm p}/N_\omega$, the free eccentricity distribution does not depend on $N_\omega$ (see Figure~\ref{fig:comparison}).

\begin{figure}
	\centering
	\includegraphics[width=\linewidth]{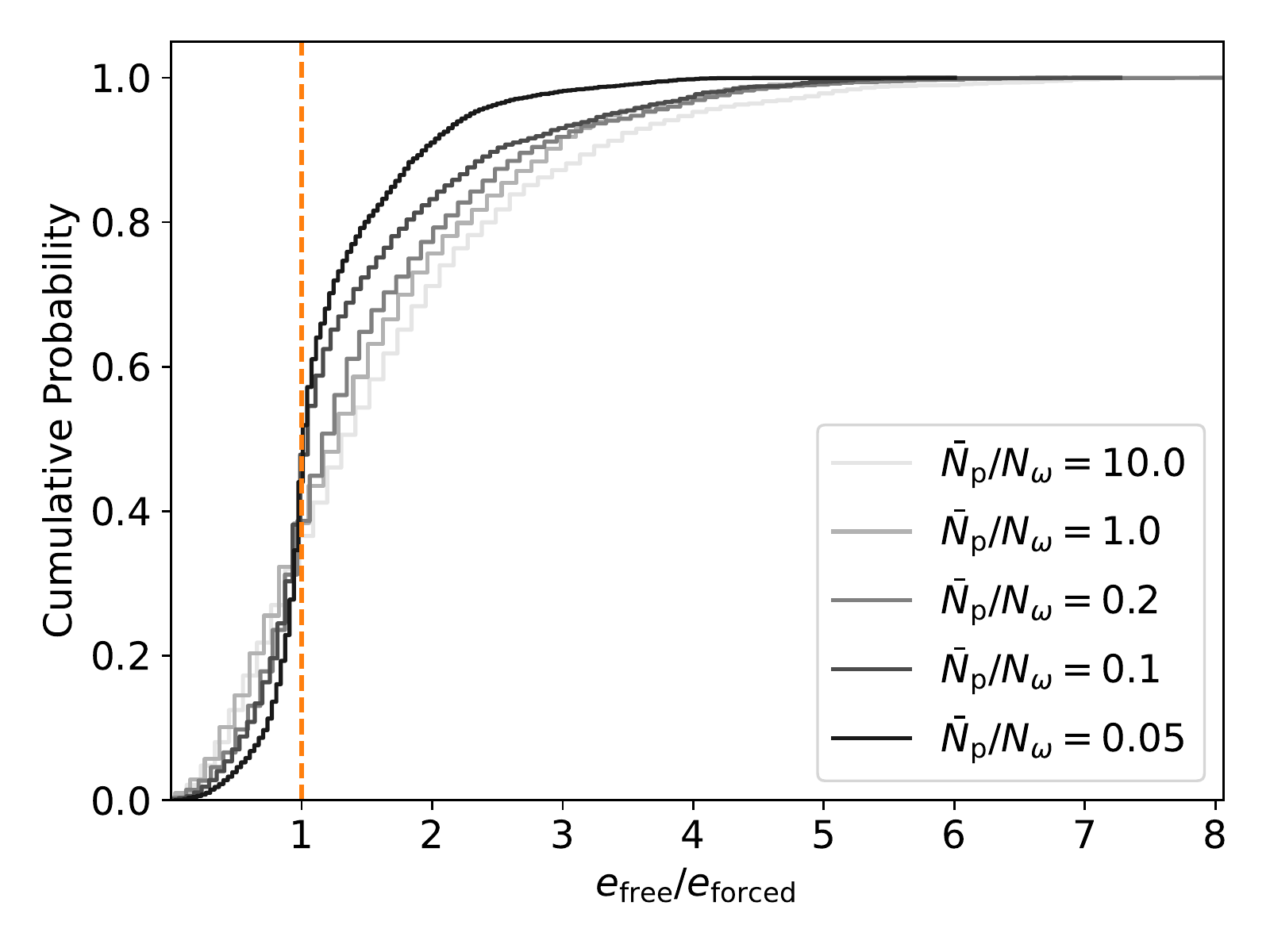}
	\caption{Cumulative distribution of the final free eccentricity of the debris belt after the planet-planet scattering process (Section~\ref{sec:randomgrowth_numerical}), with $\bar{N}_{\rm p} = 1000$ and different values of $N_\omega$. Between 30\% and 50\% of the walks have a free eccentricity less than the forced eccentricity (orange vertical line), so that the resulting debris belt would be narrower than in the standard picture where the planet attains its eccentricity instantaneously. }\label{fig:efree_randomgrowth}
\end{figure}

Since most of the walks have a length close to $\bar{N}_{\rm p}$, the overall distribution of free eccentricities depend mostly on $\bar{N}_{\rm p}/N_\omega$. Figure~\ref{fig:efree_randomgrowth} shows the cumulative distribution of free eccentricities for different $\bar{N}_{\rm p}/N_\omega$. We find that although $\langle e_{\rm free}^2 \rangle$ increases as $\bar{N}_{\rm p}/N_\omega$ increases beyond unity, a significant proportion (about $40\%$) of the random walks still lead to $e_{\rm free} \lesssim e_{\rm forced}$. This is our key result, suggesting that if the perturbing planet attains its eccentricity through random walks associated with planet-planet scatterings, the planetesimals perturbed by the planet have a significant probability of achieving small free eccentricities (compared to $e_{\rm forced}$) and the eccentric belt can therefore maintain a small width.

\subsection{Theoretical model}
\label{sec:randomgrowth_theory}

The numerical random-walk model (Section~\ref{sec:randomgrowth_numerical}) allowed us to get some insights on how a test particle reacts to a planet with randomly evolving eccentricity. In this subsection, we solve this problem analytically by considering a continuous random-walk model, inspired by the approach of \cite{puStrongScatteringsCold2021}.

In the continuous random-walk model, the initial and final planet eccentricities are still fixed to $0$ and $e_{\rm p, f}$, respectively. We define $t_{\rm p}$ the total time duration of a walk, and $\sigma_{\rm p}^2$ its diffusion constant (note that it has the unit of frequency). The random variable $\mathcal{E}_{\rm p}(t)$ then depends on time in the following way:
\begin{align}
	&\langle \mathcal{E}_{\rm p}(t) \rangle  = \frac{t}{t_{\rm p}} e_{\rm p, f},\label{eq:average}\\
	&\langle \mathcal{E}_{\rm p}(t) \mathcal{E}_{\rm p}^*(s)\rangle -  \langle \mathcal{E}_{\rm p}(t) \rangle \langle \mathcal{E}_{\rm p}^*(s)\rangle  = 2 \min(t, s) \left[1-\frac{\max(t, s)}{t_{\rm p}}\right] \sigma_{\rm p}^2. \label{eq:covariance}
\end{align} 

Assuming an initially zero eccentricity for the test particle, equation~\eqref{eq:solution} then becomes
\begin{align}
	\langle \mathcal{E}(t) \rangle &{}= \int_{0}^t \langle \mathcal{E}_{\rm p}(t') \rangle \left[-i\nu \exp i \omega (t-t') \right] \dd{t'}\\
	&{}= i e_\mathrm{forced} \frac{\exp(i\omega t) - i\omega t -1 }{\omega t_{\rm p}},
\end{align}
where the forced eccentricity is given by $e_{\rm forced} = (\nu/\omega) e_{\rm p,f}$ (see equation~\ref{eq:eforced_random}). Moreover, at $t = t_{\rm p}$, we have 
\begin{align}
	\langle |\mathcal{E}(t_{\rm p})|^2 \rangle &{}= \int_{0}^{t_{\rm p}} \int_{0}^{t_{\rm p}} \langle \mathcal{E}_{\rm p}(u) \mathcal{E}^*_{\rm p}(s)\rangle \left[-i\nu\exp i \omega (s-u) \right] \dd{s} \dd{u} \\
	&{}= 2 \sigma_{\rm p}^2 \left(\frac{\nu}{\omega}\right)^2 t_{\rm p} \left(1 - \frac{2 - 2\cos\omega t_{\rm p}}{\omega^2 t_{\rm p}^2}\right) + \langle |\mathcal{E}(t_{\rm p})| \rangle^2.
\end{align}

The complex free eccentricity at the end of the random walk is given by
\begin{equation}
	\mathcal{E}_{\rm free} = \mathcal{E}(t_{\rm p}) - e_{\rm forced}.
\end{equation}
The mean square of the free eccentricity is thus:
\begin{align}
	\langle |\mathcal{E}_{\rm free}|^2 \rangle ={}& \langle |\mathcal{E}(t_{\rm p})|^2 \rangle - 2 \mathrm{Re}\langle e_{\rm forced}\mathcal{E}(t_{\rm p}) \rangle + e_{\rm forced}^2\nonumber \\
	={}& e_\mathrm{forced}^2 \frac{2 + 2y x (x^2 - 2) - 2(1 - 2 y x)\cos x}{x^2}, \label{eq:efreerms}
\end{align}
where
\begin{align}
	&x \equiv \omega t_{\rm p}, \label{eq:x}\\
	&y \equiv \frac{\sigma_{\rm p}^2}{e_{\rm p, f}^2 \omega}\label{eq:y}.
\end{align}
The quantity $x$ measures the degree of adiabaticity of the planet eccentricity growth compared to the precession timescale. If the growth is fast, i.e. $x \ll 1$, then 
\begin{equation}
	\langle |\mathcal{E}_{\rm free}|^2 \rangle \sim e_\mathrm{forced}^2.
\end{equation}
The average $|\mathcal{E}_{\rm free}|^2$ can get lower than $e_{\rm forced}^2$ for $x\lesssim 1$ for $y^{-1} > 2x$. For $y^{-1} \gg 2x$, we recover the result of the linear planet eccentricity growth described in Section~\ref{sec:eccentricitygrowth} (equation~\ref{eq:efree_egrowth}). On the other hand, if the growth is adiabatic, i.e. $x \gg 1$, then
\begin{equation}
	\langle |\mathcal{E}_{\rm free}|^2 \rangle \sim 2 e_\mathrm{forced}^2 x y \propto e_\mathrm{forced}^2~t_{\rm p}.
\end{equation}

We note that the $x$ and $y$ parametrization is analogous to the $\bar{N}_{\rm p}$ and ${N_{\rm p}}/{N_\omega}$ description introduced in Section~\ref{sec:randomgrowth_numerical}, with the correspondence
\begin{align}
	&x = 2\upi \frac{N_{\rm p}}{N_\omega},\\
	&y = \frac{1}{4\upi} \frac{N_\omega}{\bar{N}_{\rm p}}.
\end{align}
Note that $2xy = 1$ is equivalent to $N_{\rm p} = \bar{N}_{\rm p}$. We can thus directly compare the numerical model of Section~\ref{sec:randomgrowth_numerical} and our analytical result (equation~\ref{eq:efreerms}) in Figure~\ref{fig:efree_randomgrowthvsstepstheory}. Although we manage to capture the average free eccentricity of the test particle in the continuous random-walk model, equation~\eqref{eq:efreerms} does not give any information on the spread of free eccentricities. As shown in Section~\ref{sec:randomgrowth_numerical}, a significant proportion of the free eccentricities can be lower than the forced eccentricity even as their average is high.

\section{Summary and Discussion}
\label{sec:conclusion}

\subsection{Summary}

Planetesimals can be shaped into an eccentric debris belt by a nearby eccentric planet companion. The eccentricity of each planetesimal oscillates around an equilibrium value, the so-called forced eccentricity $e_{\rm forced}$. The period of this variation is the precession period (equation~\ref{eq:tw}), and its amplitude is the free eccentricity $e_{\rm free}$. The minimum width of the belt is $\Delta r = 2 \bar{a} e_{\rm free}$, where $\bar{a}$ is the mean semi-major axis of the belt. In the standard picture, where the planet is born eccentric, the free eccentricity of the planetesimals is equal to the forced eccentricity (Section~\ref{sec:classical}). However, observations of at least three narrow eccentric debris belts suggest that $e_{\rm free}$ can be much smaller than $e_{\rm forced}$ (see Section~\ref{sec:introduction}). In this paper, we have studied two physical processes that can reduce the free eccentricity of a debris belt: (i) interaction of the planet and/or planetesimals with the protoplanetary discs; (ii) stochastic growth of the planet eccentricity through planet-planet scatterings. 

In Section~\ref{sec:exploration}, we lay down the main equations for the evolution of the planetesimals eccentricity driven by an evolving planet and demonstrate the mathematical validity of our hypotheses. We show that the free eccentricity can effectively be reduced if the planetesimals are subjected to an eccentricity damping force (Section~\ref{sec:eccentricitydamping}), or if the planet eccentricity growth is gradual (Section~\ref{sec:eccentricitygrowth}). In the first case, the force has to act for a duration comparable to the damping timescale. We also point out that if the damping is too strong, then the equilibrium eccentricity is shifted and the free eccentricity may still be large when the force stops acting. In the second case, we demonstrate that the free eccentricity is strongly reduced if the planet eccentricity reaches its final value slower than the precession timescale of the planetesimals (Figure~\ref{fig:efree}).

In Section~\ref{sec:ppdisc}, we explore these mechanisms at the beginning of the system's life, when planet and planetesimals are embedded in the protoplanetary disc. If the planet eccentricity grows due to interactions with the gas disc, a narrow eccentric debris belt could be produced. Alternatively, the gas environment can damp the free eccentricity of the planetesimals (Figure~\ref{fig:efree_friction}). In both cases however, if the separation of the belt from the host star is too large or is the planet is not very close to the belt, then the protoplanetary disc lifetime is not long enough (as compared to the precession period or the damping timescale) to induce a significant effect.

In Section~\ref{sec:ppscattering}, we study the effect of an instability phase between two planets leading to the ejection of one and the eccentricity excitation of the other. This excitation can be modelled as a random walk in the complex eccentricity plane. We combine numerical and analytical methods to study the properties of this statistical process, and find that it can lead to a wide range of free eccentricities in the neighbouring planetesimal belt. In around $40\%$ of the cases (Figure~\ref{fig:efree_randomgrowth}), the free eccentricity is smaller than the forced eccentricity, making this process a reliable candidate to account for the observed narrow widths of eccentric debris belts.

We summarize here our key points:
\begin{itemize}
	\item The shape of debris belts carry information on the dynamical history of the planet companion shaping it, and on the initial conditions of the system;
	\item Interaction with the protoplanetary disc can limit free eccentricities of the planetesimals (and thus the width of the belt), either through a friction force acting on the planetesimals, or by inducing a slow growth of the planet eccentricity. This requires the disc lifetime to be larger than the eccentricity damping timescale or the precession period, respectively;
	\item Alternatively, the stochastic eccentricity growth associated with planet-planet scattering has a significant chance to reduce the width of a debris belt compared to the often-assumed instantaneous growth. The outcome of the process is probabilistic and could also lead to increased belt widths.
\end{itemize}

\subsection{Discussion}

We argue that the processes described in our paper could adequately account for the narrow widths of planetesimal belts. However, our demonstration uses idealized semi-analytical models, which naturally have some limitations. Our initial conditions suppose that one and only one planetary companion is directly responsible for the shape of the belt (this companion can be perturbed by a gaseous disc or an other planet, but it is the only body interacting with the planetesimals). Although it is a reasonable hypothesis when the system only harbours one giant planet, it may not be the case when several giant planets co-exist (as in our Solar System).

We are also aware of the limitations of modelling the friction effect in the protoplanetary disc with a simple non-linear force, and the large uncertainties in the gas disc properties (e.g. density and lifetime) limit our conclusion. Thus, the result of Section~\ref{sec:ppdisc} should be considered as a proof of concept that helps to identify the important quantities at play. The planet-planet scattering process described in Section~\ref{sec:ppscattering} also uses some simplifying assumptions. Treating the evolution of the complex eccentricity of the planet as a pure random walk is somewhat idealized, and the assumption that the interaction between the planet and planetesimals remain secular at all time is reasonable only if they are sufficiently separated.

In our study, we have neglected non-gravitational processes that could impact the spatial distribution of dust---we assume that they perfectly trace their parent planetesimals in the belt. Further observations, in different wavelengths, will allow us to test whether narrow eccentric debris belts are common or are the result of rare conditions of formation and evolution. The detection of the planet companions will help to confirm the relation between the debris shape and planetary architecture. The companion and debris disc characterization by direct imaging instruments, ALMA and JWST will be valuable in this regard.

\section*{Acknowledgements}

We thank the referee for their useful comments that have improved this paper. This work has been supported in part by the NSF grant AST-17152 and NASA grant 80NSSC19K0444. We made use of the \textsc{python} libraries \textsc{NumPy} \citep{harrisArrayProgrammingNumPy2020}, \textsc{SciPy} \citep{virtanenSciPyFundamentalAlgorithms2020}, and the figures were made with \textsc{Matplotlib} \citep{hunterMatplotlib2DGraphics2007}.

\section*{Data Availability}

The codes and generated data sets used for the study of stochastic planet-planet scattering evolution (Section~\ref{sec:ppscattering}) are available at \url{https://github.com/LaRodet/EccentricDebrisBelts.git}. All other figures can be directly reproduced from the equations in the paper.



\bibliographystyle{mnras}
\bibliography{Biblio} 




\appendix

\section{Eccentricity damping from a non-linear friction force}
\label{sec:appendix_eccentricitydamping}

\subsection{Derivation of the eccentricity damping rate}

Let us write the acceleration $\vec{f}$ induced by the friction force (equation~\ref{eq:frictionforce}) as
\begin{align}
	\vec{f} \equiv{}& - \Delta v \frac{\vec{\Delta v}}{d_0}\nonumber\\
	={}& - \sqrt{v_r^2 + (v_\phi-v_\mathrm{Kep})^2} \left( \frac{v_r}{d_0} \vec{\hat{r}} + \frac{v_\phi - v_\mathrm{Kep}}{d_0} \vec{\hat{\phi}} \right),
\end{align}
with $v_\mathrm{Kep}=\sqrt{GM/r}$ the Keplerian velocity, $v_r$ the radial component of the test particle velocity, and $v_\phi$ its azimuthal component. We define the timescale $\tau_0 \equiv d_0/\sqrt{GM/a}$ and assume that it is larger than the orbital period $P$. 

The Gauss planetary equations give:
\begin{align}
	\dv{e}{t} &{}= \sqrt{\frac{a (1-e^2)}{GM}} \left[f_r \sin \phi + f_\phi(\cos \phi + \cos u) \right] \\
	\dv{\varpi}{t} &{}= - \sqrt{\frac{a (1-e^2)}{GMe^2}}f_r \cos \phi\nonumber\\
	&{}\quad + \left(\frac{r}{\sqrt{GMa\sqrt{1-e^2}}}+\sqrt{\frac{a(1-e^2)}{GM}}\right) f_\phi \sin \phi, 
\end{align}
where $u$ and $\phi$ are respectively the eccentric and true anomaly.

To the leading order in $e$, $f_r$ and $f_\phi$ are given by:
\begin{align}
	f_r &{} = -e^2 \frac{GM}{a d_0} \sin u \sqrt{1 - \frac{3}{4}\cos^2u} + O(e^3)\\
	f_\phi &{} = -e^2 \frac{GM}{a d_0} \frac{\cos u}{2} \sqrt{1 - \frac{3}{4}\cos^2u} + O(e^3).
\end{align}
These give
\begin{align}
	\dv{e}{t} &{}= -\frac{e^2}{\tau_{0}} \sqrt{1 - \frac{3}{4}\cos^2u} + O(e^3).
\end{align}
Averaging over one orbital period, we have
\begin{align}
	\langle\dv{e}{t} \rangle &{}= -\frac{e^2}{\tau_{0}} \int_0^{2\upi} \dd{u} \frac{r}{2\upi a} \sqrt{1 - \frac{3}{4}\cos^2u} + O(e^3)\nonumber\\
		&{}= -C_e \frac{e^2}{\tau_{0}}  + O(e^3),
\end{align}
where $C_e = E(-3)/\upi \simeq 1$.

Similarly, we can derive $\langle\dd{\varpi}/\dd{t}\rangle$ and find that it vanishes at least at order $2$ in $e$.

\subsection{Equilibrium eccentricity}

The equilibrium eccentricity of a test particle experiencing the frictional force and excitation from an inner planet is given by
\begin{equation}
	i\omega \left(\mathcal{E} - e_{\rm forced}\right)  - |\mathcal{E}| \frac{\mathcal{E}}{\tau_{0}} = 0.
\end{equation}
This equation can be solved analytically and has two real solutions:
\begin{align}
	\mathcal{E}_{\rm eq} = {}&\tau_{0}\omega \frac{-\tau_{0}\omega + \sqrt{4 e_{\rm forced}^2+\tau_0^2\omega^2}}{2e_{\rm forced}} \pm \nonumber\\
	& i \tau_{0}\omega \Big[\frac{-3\tau_{0}\omega + \sqrt{4 e_{\rm forced}^2+\tau_0^2\omega^2}}{\sqrt{2}} +\nonumber\\
	& \frac{\tau_{0}^2\omega^2 (\tau_{0}\omega + \sqrt{4 e_{\rm forced}^2+\tau_0^2\omega^2})}{\sqrt{2}e^2_{\rm forced}}\Big].\label{eq:eeq}
\end{align}
The shift between $\mathcal{E}_{\rm eq}$ and the standard forced eccentricity $e_{\rm forced} = (\nu/\omega) e_{\rm p}$ is
\begin{align}
	&\sqrt{e^2_{\rm forced}+ \frac{\tau_{0}\omega}{2}\left(\tau_{0}\omega - \sqrt{4 e_{\rm forced}^2+\tau_0^2\omega^2}\right)}\nonumber\\
	&{} = \frac{e_{\rm forced}}{\tau_{0}\omega} + O(e_{\rm forced}^4).
\end{align}
The shift will become significant if $\tau_{0} \omega \sim e_{\rm forced} < 1$.

\section{Additional information on the random walk numerical modelling}
\label{sec:randomwalk_method}

\begin{figure}
	\centering
	\includegraphics[width=\linewidth]{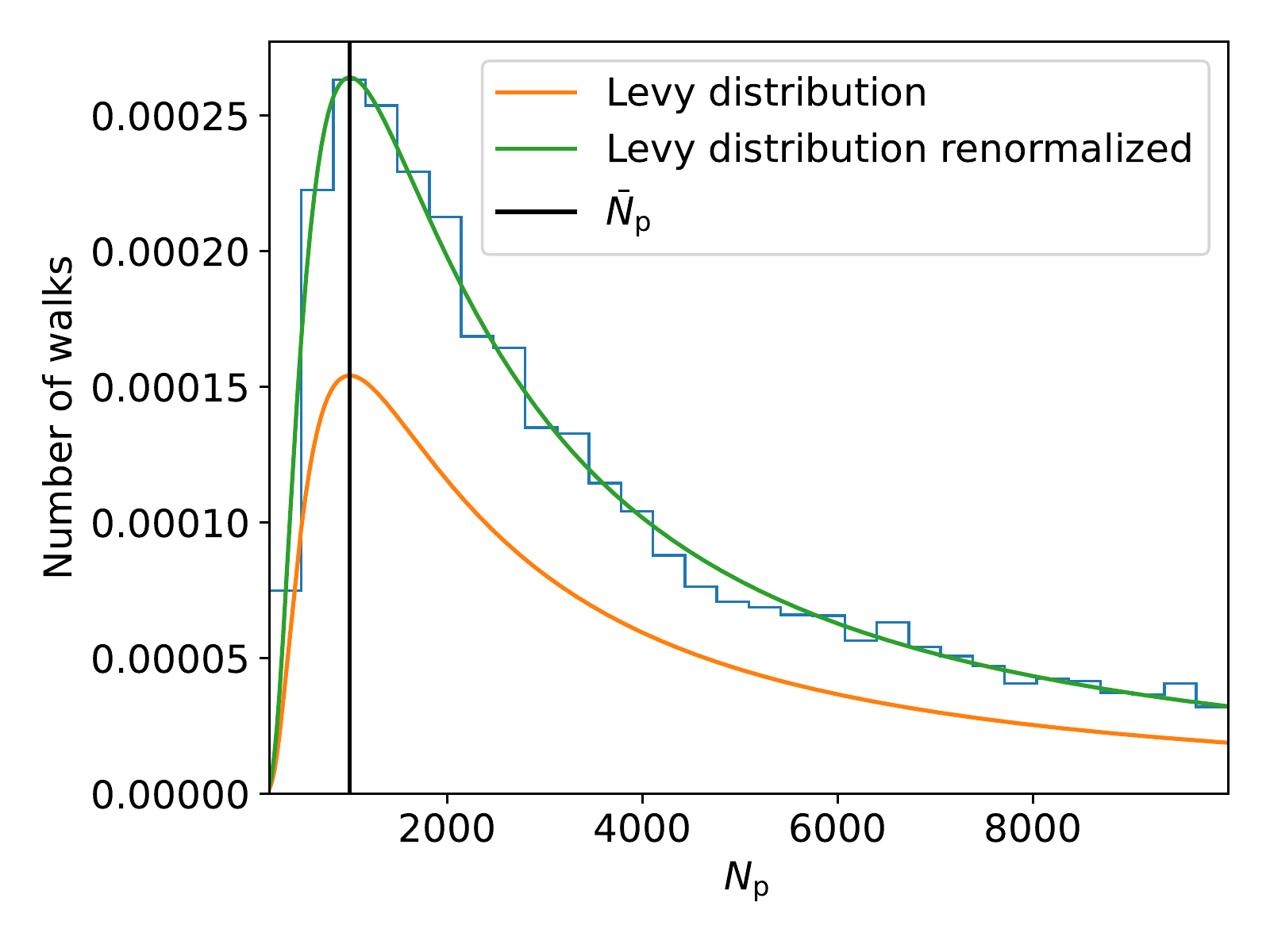}
	\caption{Distribution of random walk lengths (the number of kicks) among our sample (blue), compared to a Lévy distribution peaking at $\bar{N}_{\rm p} = 1000$ (orange). We restrict our walks to length less than $10~\bar{N}_{\rm p}$. Renormalizing the theoretical Lévy distribution to take into fount this cut-off, we get the green curve. }\label{fig:distribution}
\end{figure}

We want to sample random walks of length $n$ with $n$ following a Lévy distribution:
\begin{equation}
	f(n) = \frac{b}{\sqrt{2\upi n^3}} \exp\left(-\frac{b^2}{2n}\right),
\end{equation}
where $b = \sqrt{3 \bar{N}_{\rm p}}$. The function $f$ is maximal at $n = \bar{N}_{\rm p}$. This is equivalent to sampling uniformly a parameter $p$ such that
\begin{equation}
	\dd{p} = f(n) \dd{n}. \label{eq:dp}
\end{equation}
Let us define $p$ as:
\begin{equation}
	p = 1 - \int_0^{\frac{b}{\sqrt{2n}}} \exp(x) \dd{x} = \mathrm{erfc} \left(\frac{b}{\sqrt{2n}}\right) \label{eq:p},
\end{equation}
where erfc is the complementary error function, or equivalently
\begin{equation}
	n = \frac{b^2}{2~\mathrm{erfcinv}(p)^2}, \label{eq:n}
\end{equation}
where erfcinv is the inverse of the complementary error function. Both erfc and erfcinv are encoded into the \textsc{SciPy} Python library. The quantity $p$ as defined in equation~\eqref{eq:p} satisfies equation~\eqref{eq:dp}, so that we generate a sample of $p$ with a uniform random generator and derive the corresponding ensemble of $n$ using equation~\eqref{eq:n}. The probability distribution of the generated $n$ is shown in Figure~\ref{fig:distribution}.

In practice, since $f$ goes very slowly to $0$, we define the lower and upper limits of $n$ such that $n_{\rm min} = \sqrt{2 \bar{N}_{\rm p}}$ and $n_{\rm max} = 10\bar{N}_{\rm p}$. We sample a total of $10,000$ walks. The resulting distribution is plotted in Figure~\ref{fig:distribution}.

Once we have a distribution of random walks of size $n$, we have to generate the corresponding walks. The requirements given by Section~\ref{sec:randomgrowth_numerical} are that they begin at $0$, end at $\mathcal{E}_{\rm p} = e_{\rm p,f}$, with a step size $\Delta e_{\rm p}$. We relax the requirement to include walks whose final module is $e_{\rm p,f}$---there is no preferred direction in our numerical toy model. We generate walks of size $n_{\rm max}$. Within these walks, we record all sub-sequences that begins by $0$ and ends with $|\mathcal{E}_{\rm p}|$ within $\Delta e_{\rm p}/2$ of $e_{\rm p,f}$. We then perform a global rotation on the elements of the walks, so that the initial and final values are $\mathcal{E}_{\rm p} = 0$ and $\mathcal{E}_{\rm p} = e_{\rm p,f}$ respectively. We generate these walks until the entire Levi distribution is filled.

\begin{figure*}
	\centering
	\includegraphics[width=0.45\linewidth]{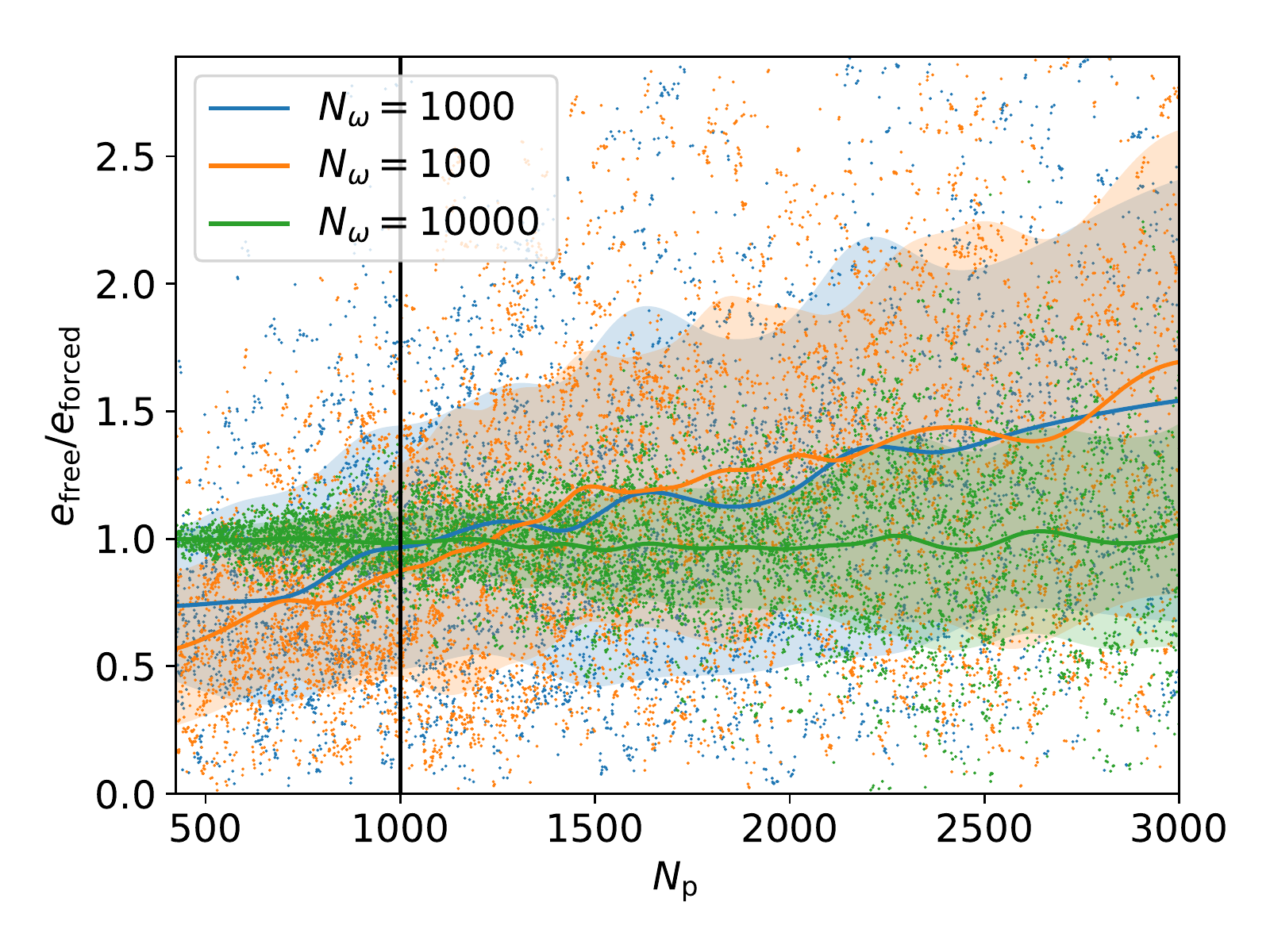}
	\includegraphics[width=0.45\linewidth]{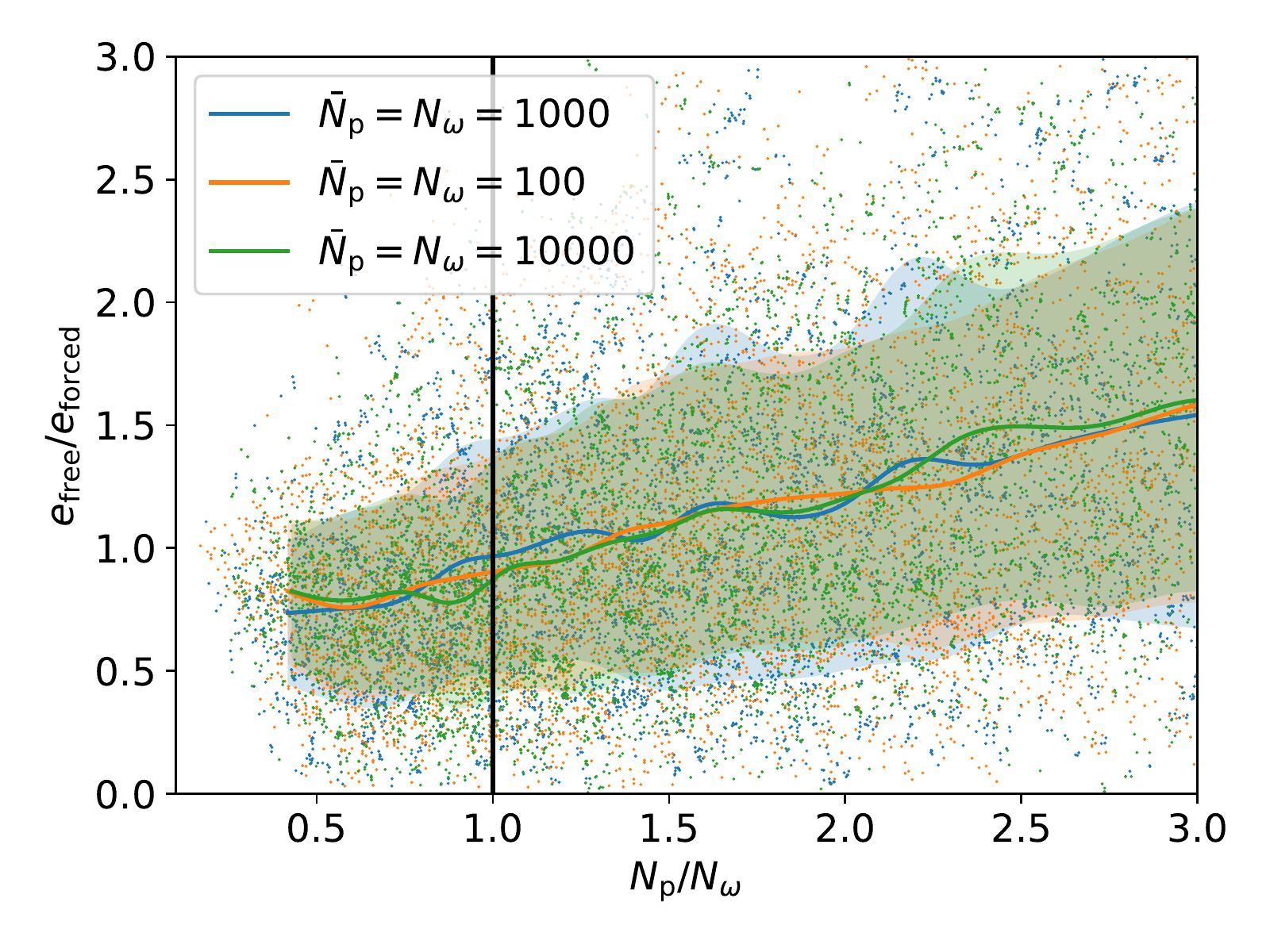}
	\caption{Final free eccentricity of the debris belt as a function of the number of kicks assuming a 2D random eccentricity growth of the planet. Left panel: $\bar{N}_{\rm p} = 1000$, and $N_\omega = 100, 1000$ and $10,000$. Right panel: $\bar{N}_{\rm p} = N_\omega$, and $N_\omega = 100, 1000$ and $10,000$. Data with similar $\bar{N}_{\rm p}/N_{\omega}$ (i.e. $y$, equation~\ref{eq:y}) exhibit the same dependency in $N_{\rm p}/N_{\omega}$(i.e. $x$, equation~\ref{eq:x}), as discussed in Section~\ref{sec:randomgrowth_theory}.}\label{fig:comparison}
\end{figure*}


\bsp	
\label{lastpage}
\end{document}